\documentclass[pra,twocolumn,superscriptaddress,10pt,noshowpacs]{revtex4}
\usepackage[english]{babel}
\usepackage[T1]{fontenc}
\usepackage[utf8]{inputenc}
\usepackage{graphicx,epstopdf}
\usepackage{amssymb}
\usepackage{amsmath}

\usepackage{amsfonts}
\usepackage{bbm}
\usepackage{color}
\usepackage{latexsym}
\usepackage{caption}
\usepackage{subcaption}
\usepackage{times,txfonts}
\usepackage{color,soul}
\usepackage{listings}
\usepackage{bbold}

\newcommand{\beq}{\begin{equation}}
\newcommand{\eeq}{\end{equation}}
\newcommand{\bea}{\begin{eqnarray}}
\newcommand{\eea}{\end{eqnarray}}
\begin{document}

\title{Strain effects on the electronic properties of a graphene wormhole}

\author{J. E. G. Silva}
\email{euclides@fisica.ufc.br}
\affiliation{Universidade Federal do Cear\'a, Departamento de F\'{i}sica, 60455-760, Fortaleza, CE, Brazil}

\author{\"{O}. Ye\c{s}ilta\c{s}}
\email{yesiltas@gazi.edu.tr}
\affiliation{Department of Physics, Faculty of Science, Gazi University, 06500 Ankara, Turkey}

\author{J. Furtado}
\email{job.furtado@ufca.edu.br}
\affiliation{Universidade Federal do Cariri, Centro de Ci\^encias e Tecnologia, 63048-080, Juazeiro do Norte, CE, Brazil}
\affiliation{Department of Physics, Faculty of Science, Gazi University, 06500 Ankara, Turkey}

\author{A. A. Ara\'{u}jo Filho}
\email{dilto@fisica.ufc.br}
\affiliation{Departamento de Física Teórica and IFIC, Centro Mixto Universidad de Valencia--CSIC. Universidad de Valencia, Burjassot-46100, Valencia, Spain}
\affiliation{Departamento de Física, Universidade Federal da Paraíba, Caixa Postal 5008, 58051-970, João Pessoa, Paraíba, Brazil}

\date{\today}

\begin{abstract}
In this work, we explore the strain and curvature effects on the electronic properties of a curved graphene structure, called the graphene wormhole. The electron dynamics is described by a massless Dirac fermion containing position--dependent Fermi velocity. In addition, the strain produces a pseudo--magnetic vector potential to the geometric coupling. For an isotropic strain tensor, the decoupled components of the spinor field exhibit a supersymmetric (SUSY) potential, depending on the centrifugal term and the external magnetic field only. In the absence of an external magnetic field, the strain yields an exponentially damped amplitude, whereas the curvature leads to a power--law damping of the wave function. The spin--curvature coupling breaks the chiral symmetry between the upper and the lower spinor component, which leads to the increasing of the wave function on either upper or lower region of the wormhole, i.e., depending on the spin number. By adding a uniform magnetic field, the effective potential exhibits an asymptotic quadratic profile and a spin--curvature barrier near the throat. As a result, the bound states (Landau levels) are confined around the wormhole throat showing an asymmetric and spin--dependent profile.
\end{abstract}


\maketitle

\section{Introduction}
\label{introduction}

Two dimensional materials, such as graphene \cite{geim}, silicene \cite{silicene} and phosphorene \cite{phosphorene}, have been the subject of intense investigation due to their outstanding properties.
Beyond the remarkable mechanical \cite{katsnelson} and electronic properties \cite{Novoselov2004,electronic}, graphene can also be seen as a table--top laboratory for relativistic physics. Indeed, since the conduction electrons are effectively described as massless Dirac fermions, relativistic effects such as zitterbewegun \cite{zitter}, Klein tunneling \cite{klein} and atomic collapse \cite{collapse} have been observed. Since the graphene layer can assume a curved shape, the curvature effects might lead to new interesting relativistic effects, such as the Hawking--Unruh effect \cite{hawking,hawking2}.

The study of a Dirac fermion confined into a two dimensional surface was initially addressed in Ref.\cite{BJ} and further developments were provided afterwards \cite{diracsurface,diracsurface2,diracsphere}. For a relativistic fermion intrinsically living on a curved surface, a physical realization was found for conducting electrons on two dimensional carbon--based structures, as the fullerenes \cite{diracintrinsic,diracintrinsic1}, carbon nanotubes \cite{saito} and graphitic cones \cite{cone}.
In graphene, the massless Dirac equation in curved spaces was studied in a variety of shapes, such as the localized gaussian bump \cite{contijo}, the cone \cite{furtado}, a helical graphene ribbon \cite{atanasov, watanabe}, a corrugated plane \cite{corrugated}, a M\"{o}bius ring \cite{mobius1, mobius2}, a torus \cite{ozlem} among others.
The surface curvature produces a spin--curvature coupling which leads to a geometric Aharonov--Bohm--like effect \cite{geometricphase}, a modified spin--orbit coupling \cite{geometricsoc, geometricsoc2} and a geometric spin--Hall effect \cite{geometricmonopole}.

In addition to the curvature, the deformations of the graphene layer modify the effective Dirac fermion dynamics as well, producing the so--called pseudo--magnetic fields \cite{ribbons}. This vector potential stems from the strain tensor defined by the deformations of the graphene layer and the pseudo--magnetic term comes from the coupling to the Dirac fermion; it is similar to the minimal coupling to a magnetic field \cite{gaugestrain}. The strain applied to graphene can mimic a strong magnetic field \cite{strainstrong} and leads to important applications \cite{strainappli}. From the strain tensor, an effective Hamiltonian for the Dirac fermion was derived using the tight--binding approach \cite{vozmediano}, containing an anisotropic and position--dependent Fermi velocity and a pseudo--magnetic strain vector in the continuum limit. Since the strain may contain both in--plane and out--of--plane components, the effective Dirac Hamiltonian was extended in order to encompass all the stretching and bending effects \cite{vozmediano2}. In addition, the quantum field interaction of the effective Dirac fermion and the strain was discussed in \cite{sinner,gaugegrapheneqft}.

An interesting curved graphene structure is the so--called graphene wormhole, where two flat graphene layers are connected by a carbon nanotube \cite{wormhole}. Since its shape (cylinder) has a non--vanishing mean curvature, the discontinuity of the curvature at the graphene--nanotube junction leads to modifications of the energy spectrum and the possibility of localized states close to it \cite{graphenejunction}. Although Dirac fermions on the upper and lower layers are free states (non--normalizable), the curvature of the nanotube allows the existence of normalizable zero--modes confined at the radius of the wormhole \cite{picak,wormhole3}. In order to avoid the discontinuity at the junction, a smooth graphene wormhole was proposed considering a continuum and asymptotic flat catenoid surface \cite{dandoloff, euclides}. The negative curvature of the catenoid leads to a repulsive spin--curvature coupling near the wormhole throat, allowing only the zero--mode as a localized state around the throat \cite{wormhole4,ozlem2,wormhole5}.

In this work, we consider the effects of the curvature and the strain on the effective Dirac fermion living in a catenoid--shaped graphene wormhole. We extend the effective Hamiltonian obtained in \cite{vozmediano} to a curved surface, introducing the usual spin--connection coupling. We explore the different effects driven by the curvature, isotropic strain and an external magnetic field. Since the surface is asymptotically flat, the lattice deformation which produces the curvature and strain should be concentrated around the throat. The strain leads to a vector potential along the surface meridian, whereas the spin--curvature coupling points in the angular direction. Moreover, the strain vector potential provides an exponential damping of the wave function, whereas the curvature leads to a power--law decay. By adopting the so-called supersymmetric quantum mechanic-like approach \cite{ozlem2}, the spinor components exhibit a chiral symmetry breaking. Indeed, the upper component has its probability density enhanced near the wormhole throat in the upper layer, whereas the lower component is enhanced in the lower layer. The ground state zero mode also exhibits this chiral behaviour, since it is exponentially damped either in the upper or in the lower layer depending on the total angular momentum. By applying a uniform magnetic field, the Landau levels are also modified by the curved geometry and strain, leading to asymmetric localized states near the throat.

This work is organized as follows. In  section (\ref{section2}), we provide a brief review of the geometry of the catenoid-shaped graphene wormhole. In section (\ref{section3}) we present the effective Hamiltonian containing the strain, curvature and external magnetic field interactions. The section (\ref{section4})
is devoted to the symmetries of the effective Hamiltonian and in the section (\ref{section5}) we employ the SUSY-QM approach in order to investigate the effects of each interaction. Finally, additional discussion and perspectives are given in the section (\ref{section6}).

\section{Graphene wormhole geometry}
\label{section2}

\begin{figure}[h!]
    \centering
    \includegraphics[scale=0.3]{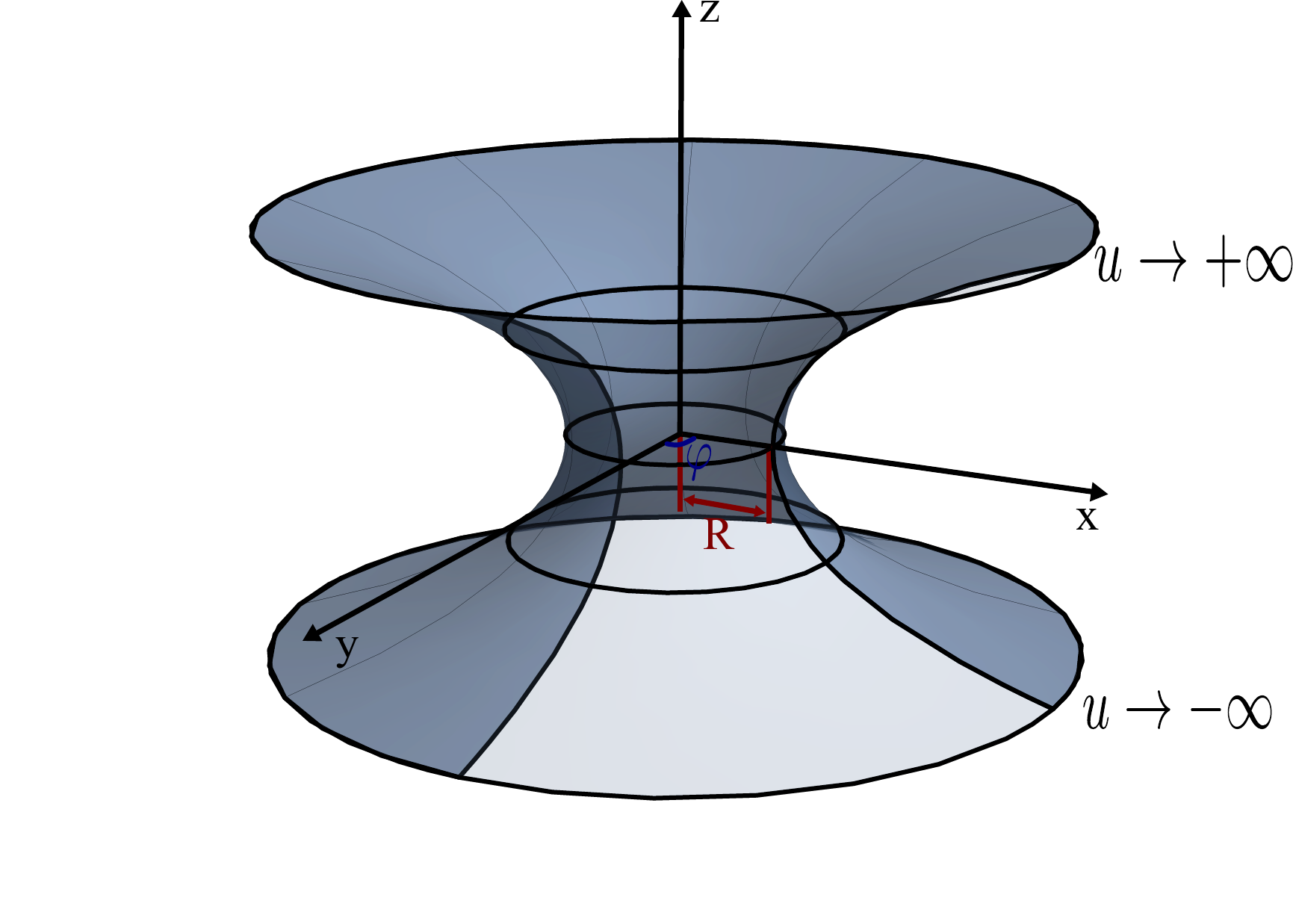}
    \caption{Graphene wormhole geometry. The meridian coordinate $u$ connects the lower to the upper asymptotic flat regions.}
    \label{fig1}
\end{figure}

In this section, we define the surface of the graphene wormhole and describe some of its most important properties. We consider a smooth surface connecting an upper to the lower layer (flat planes). For this purpose, we choose a catenoid shaped surface. In other words, the catenoid surface can be described in coordinates by \cite{euclides,ozlem2}
\begin{equation}
\label{coordinates}
    \vec{r}(u,\phi)=\sqrt{R^2+u^2}\left(\cos\phi\hat{i}+\sin\phi\hat{j}\right)+R\sinh^{-1}\left(\frac{u}{R}\right)\hat{k},
\end{equation}
where $R$ is the throat radius, $-\infty <u<\infty$ describes the meridian coordinate and $\varphi$ is the parallel coordinate $\varphi\in[0,2\pi)$, as shown in fig.\ref{fig1}. 
The tangent vectors are given by
\begin{eqnarray}
\vec{e}_1 &=& \frac{\partial \vec{r}}{\partial u}=\frac{1}{\sqrt{u^2 + R^2}}(u\hat{r}+R\hat{k})\\
\vec{e}_2 &=&\frac{\partial\vec{r}}{\partial\phi}=\sqrt{u^2 + R^2}\hat{e}_2,
\end{eqnarray}
where $\hat{e}_2=\cos\phi \hat{i}+\sin\phi\hat{j}$ is the unit vector along the $\phi$ direction. 
From the tangent vectors $(\hat{e}_1 , \hat{e}_2)$, we can define the surface induced metric $g_{ij}=\vec{e}_i \cdot \vec{e}_j$. In $(u,\phi)$ coordinates, the surface metric takes the form $g_{ij}=diag(1,(R^2 + u^2))$. Thus, the $2+1$ spacetime interval has the form \cite{euclides}
\begin{equation}
    \label{catenoidmetric}
    \mathrm{d}s^2 = \mathrm{d}t^2 - \mathrm{d}u^2-(R^2+u^2)\mathrm{d}\phi^2,
\end{equation}
where we adopt the $(+,-,-)$ spacetime metric signature convention.
Note that the line element in eq.(\ref{catenoidmetric}) is invariant under time--translations and rotations with respect to the $\mathrm{z}$ axis (axissymmetric). 

\begin{figure}
    \centering
\includegraphics[scale=0.45]{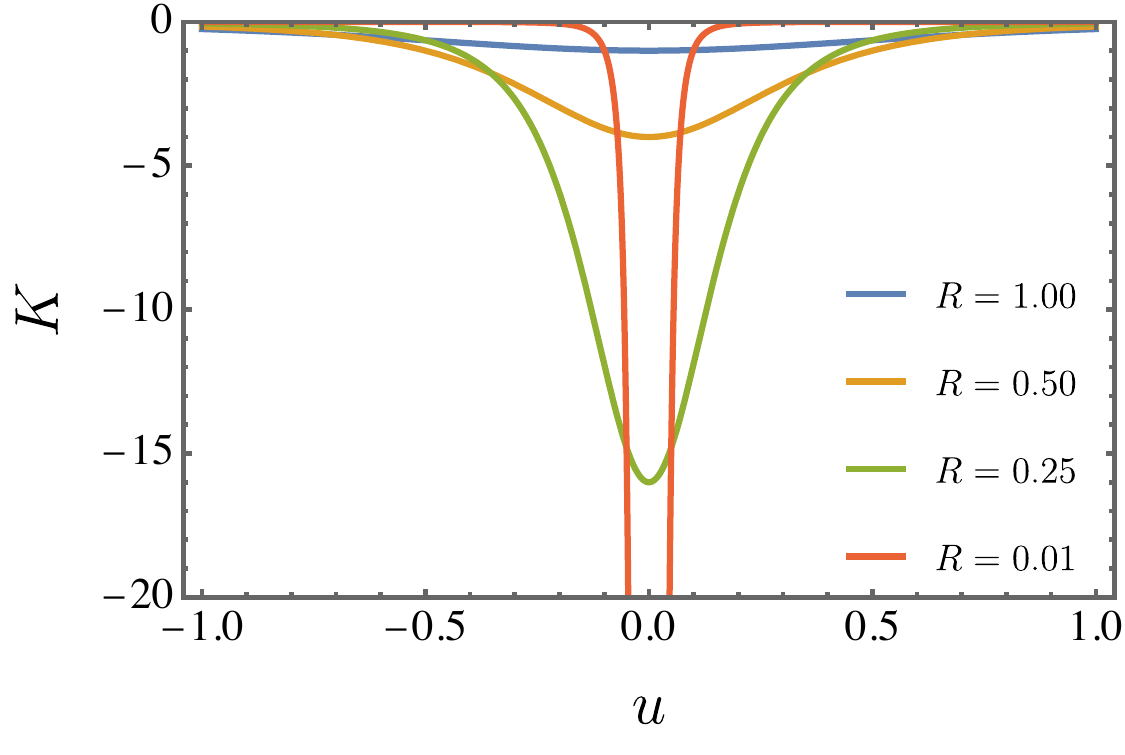}
    \caption{The Gaussian curvature $K(u)$ of the graphene wormhole. The curvature is smooth and concentrated around the throat of the wormhole. For $R\rightarrow 0$, the curvature tends to a delta--like function.}
    \label{K}
\end{figure}

Let us now obtain the the main geometric quantities for the electron dynamics, namely, the dreinbeins, connections and curvature. The dreinbeins are related to the spacetime metric by the relation 
\begin{equation}
\label{vielbeins}
g_{\mu\nu}=e^{a}_{\mu}e^{b}_{\nu}\eta_{ab}.     
\end{equation}
Thus, for the catenoid, the only non--vanishing components of the dreinbeins are
\begin{eqnarray}
\label{dreinbeins}
e^{a}_{\mu} &=& diag(1,1,\sqrt{R^2 +u^2}) 
\end{eqnarray}
Remember that they modify the Fermi velocity by turning it into a position dependent configuration. Moreover, from the dreinbeins, we can define the moving frame $\theta^{a}=e^{a}_\mu \mathrm{d}x^\mu$, where, in the catenoid, takes the form $\theta^0 = \mathrm{d}t$, $\theta^1 = \mathrm{d}u$, $\theta^2=\sqrt{R^2 + u^2}\mathrm{d}\phi$. 
From the torsion--free condition, i.e., $T^a=\mathrm{d}\theta^a+\omega^a_b\wedge \theta^b=0$, the only non--vanishing one--form connection coefficient $\omega^{a}_b = \Gamma_{cb}^{a}\theta^c$ is given by
\begin{eqnarray}
\label{connectiononeform}
        \omega^{2}_1 &=&\frac{u}{R^2 +u^2}\theta^2.
\end{eqnarray}
The curvature 2--form $R^{a}_{b}=d\omega^{a}_{b} + \omega^{a}_{c}\wedge \omega^{c}_{b}$  has only one non--vanishing component, namely $R^{2}_{1}=-\frac{R^2}{(R^2 + u^2)^2} \theta^{2}\wedge \theta^{1}$. Accordingly, the gaussian curvature $K=\delta^{ab}R_{ab}$ has the form
\begin{eqnarray}
K=-\frac{R^2}{(R^2 +u^2)^2}.
\end{eqnarray}
Here, it is important to point out that the catenoid has a negative Gaussian curvature concentrated around the throat and it vanishes in the regions far from it. In fig. \ref{K}, we display the behavior of graphene wormhole curvature $K$. Note that the surface is asymptotically flat. Thus, the effects of the curved geometry and strain on the electron should be concentrated around the origin. Furthermore, as $R\rightarrow 0$,
the curvature tends to a $\delta(r)$ function, as it is reported in the literature for a discontinuous graphene wormhole \cite{wormhole,picak,wormhole3}.

\section{Strain Hamiltonian}
\label{section3}
After a brief review of the main geometric properties of the graphene wormhole, let us now discuss the effective Hamiltonian containing the strain and curvature effects on the electron.
We follow closely the Ref.(\cite{vozmediano}) where the most general effective Hamiltonian was derived. The effective Hamiltonian for a Dirac fermion constrained to a flat surface under the influence of the strain and an external magnetic field was found in Ref. (\cite{vozmediano}). We propose a generalization of the effective Hamiltonian of Ref.(\cite{vozmediano}) in the continuum limit for curved surfaces in the form
\begin{equation}
\label{strainhamiltonian}
    \mathcal{H}_D=-i\hbar\left(v_i^j\sigma^i\partial_j + ie\sigma^{i} A_i +v_0\sigma^i \Gamma_i + v_0\sigma^i \Omega_i\right),
\end{equation}
where $v_i^j$ is a position--dependent Fermi velocity tensor defined in terms of the strain tensor $u_{ij}$ as \cite{vozmediano}
\begin{equation}
\label{velocitytensor}
    v_i^j=v_0\left[\delta_i^j-\frac{\beta}{4}(2u_i^j+\delta_i^ju_k^k)\right],
\end{equation}
and $v_0=\frac{3t_0 a}{2}$ is the undeformed Fermi velocity, 
$t$ is the hopping parameter, $a$ is the lattice constant and $\beta=|\partial \ln t/\partial \ln a|$ \cite{gaugestrain}. The definition of the strain tensor will be given in the next subsection.
Note that, when $\beta=0$, the usual constant Fermi velocity is recovered. Besides, the tensor nature of $v_{ij}$ means that the Fermi velocity depends on the direction on the surface. 

In addition, the strain on the surface also induces a new vector field, called the strain vector $\Gamma_i$, as a divergence of the velocity tensor \cite{vozmediano}. Thus, for a curved surface, it is defined as
\begin{equation}
\label{strainvectorfield}
    \Gamma_i=\frac{1}{2v_0}\nabla_jv^j_i,
\end{equation}
where the definition of the strain vector in Eq.(\ref{strainvectorfield}) is independent of the coordinate choice.

The curved Pauli matrices are defined as \cite{watanabe,mobius1,mobius2}
\begin{equation}
\label{curvedsigmas}
    \sigma^{i}=e^{i}_a \sigma^{a},
\end{equation}
where $\sigma^{a}$ are the usual flat Pauli matrices, and $e^{i}_a$ are the zweinbeins matrices which satisfy 
\begin{equation}
    g_{ij}=e^{a}_{i}e^{b}_{j}\delta_{ab}.
\end{equation}
The definition of the curved sigma matrices employed in eq.(\ref{curvedsigmas}) ensures that these matrices do not depend on the particular choice of coordinates of the surface (surface covariance). It is worthwhile to mention that the definitions of the velocity tensor in eq.(\ref{velocitytensor}) and the curved Pauli matrices in eq.(\ref{curvedsigmas}) lead to a position and direction dependent Dirac kinetic term $H_1 = v_i^j\sigma^i\partial_j$.

\begin{figure*}[ht] 
       \begin{minipage}{0.48 \linewidth}
           \includegraphics[width=\linewidth]{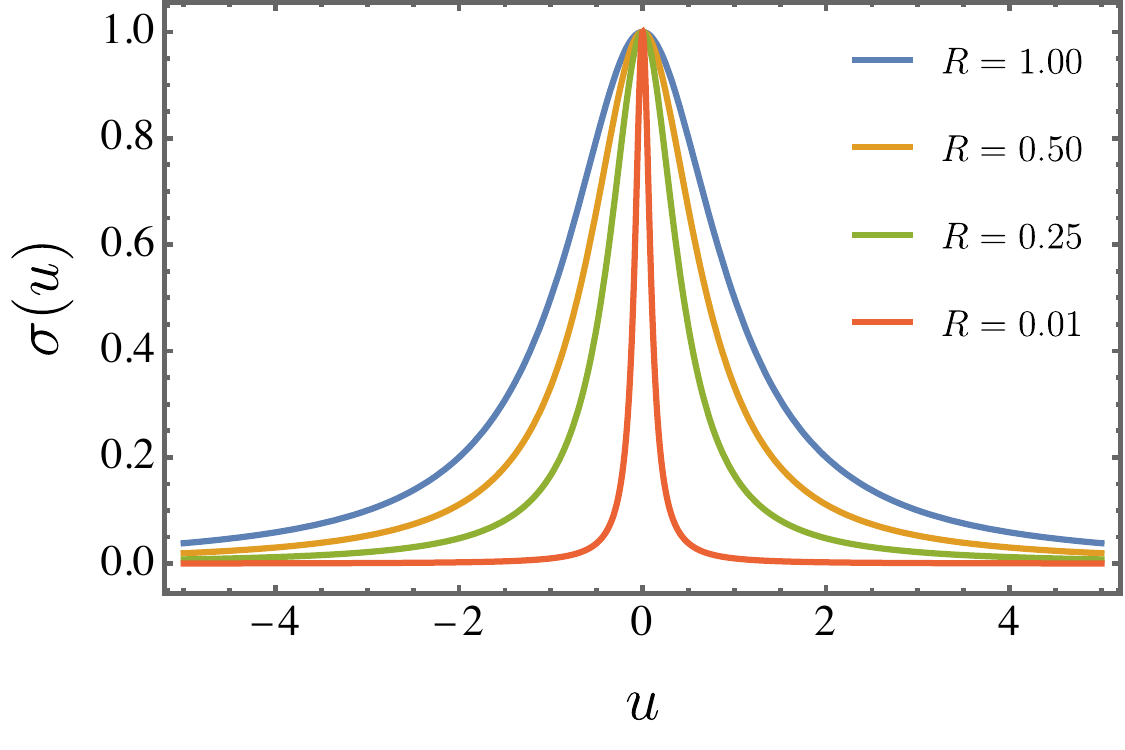}\\
           \caption{Stress function $\sigma(u)$ for}
          \label{stress}
       \end{minipage}\hfill
       \begin{minipage}{0.5 \linewidth}
           \includegraphics[width=\linewidth]{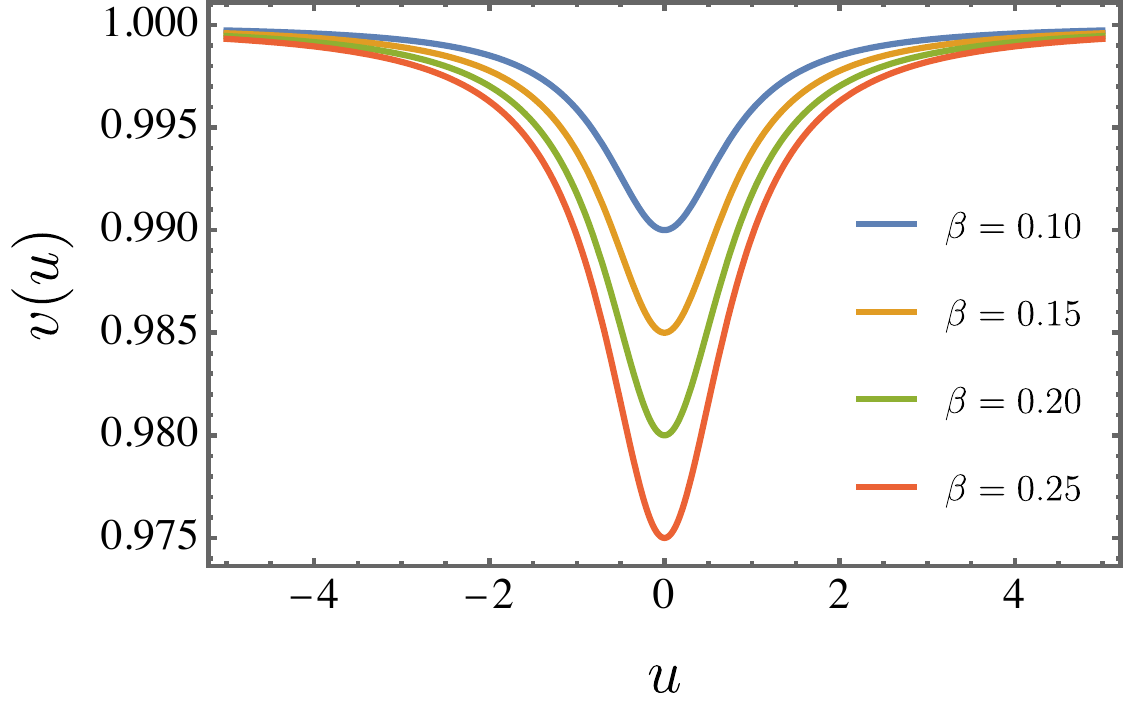}\\
           \caption{Fermi velocity $v(u)$ for $\beta=0.1$}
           \label{velocity}
       \end{minipage}
\end{figure*}

In the effective Hamiltonian exhibited in eq.(\ref{strainhamiltonian}), $A_i$ is the external magnetic potential and $\Omega_i$ is the spinor connection \cite{furtado,mobius2,ozlem2}
\begin{equation}
    \Omega_i=\frac{1}{4}\omega_{i}^{ab}\gamma_a\gamma_b.
\end{equation}
The curved $\gamma^{\mu}$ matrices are related to the flat $\gamma^{a}$ ones by the \textit{dreinbeins} $e^{a}_\mu$, i.e., $\gamma^{\mu}=e^{\mu}_a\gamma^a$. The dreinbeins are defined as
$g_{\mu\nu}=e_{\mu}^{a}e_{\nu}^{b}$. In $(2+1)$ dimension, we can adopt the following representation for the flat Dirac $\gamma^{a}$ matrices 
$\gamma_0=\sigma_3$, $\gamma_1=-i\sigma_2$ and $\gamma_2=-i\sigma_1$ \cite{mobius2,ozlem,ozlem2}. Thus, the curved Dirac matrices on the wormhole graphene surface have the form
\begin{eqnarray}
\gamma^t&=&e^t_0\gamma^0=\gamma_0,\\
\gamma^u&=&\gamma^1,\\
\gamma^\phi&=&e_2^{\phi}\gamma^2=\frac{1}{\sqrt{R^2+u^2}}\gamma^2.
\end{eqnarray}
From the connection 1--form in eq.(\ref{connectiononeform}), only $\omega^{2}_1$ is non-zero. 
Thus, the only non--vanishing component of the spinor connection $\Omega_\mu$ is
\begin{equation}
\label{spinorconnection}
    \Omega_{\varphi}=\frac{i}{2}\frac{u}{\sqrt{R^2+u^2}}\sigma_3 .
\end{equation}
Note that, since $-\infty <u<\infty$, the geometric spinor potential in eq.(\ref{spinorconnection}) is an odd function under parity. For $R=0$ or for $u\rightarrow\pm\infty$, the spinor connection becomes constant $\Omega_\phi = \frac{i}{2}\sigma_3$, as obtained in the graphitic cone \cite{furtado}.  the geometric connection is constant, as found for conic surfaces \cite{furtado}. 
It is worth mentioning that, due to the resemblance of the spinor and gauge field couplings, the spinor potential is sometimes interpreted as a kind of a pseudo--magnetic potential stemming from the curved geometry \cite{contijo}.

Furthermore, the strain produces two different potentials on the Dirac electron. The first potential, due to the strain in eq.(\ref{strainvectorfield}), is a vector potential, while the second one in eq.(\ref{spinorconnection}) is a spinorial potential depending on the surface connection. In the next subsections, we choose a particular configuration for the strain and the external magnetic field and explore the differences between these three interactions. 

\subsection{Strain tensor}
Now, let us investigate how the strain tensor $u_{ij}$ modifies the catenoid surface. In order to do it, we assume that the tensions over the surface are static and isotropic. Therefore, we consider the non--uniform isotropic stress tensor in the form
\begin{equation}
\label{strainansatz}
    \sigma_j^i=\sigma(u)\delta_j^i.
\end{equation}
Since the catenoid brige is an asymptotically flat surface, we are interested in stress tensor which vanishes away from the throat and it is finite at the origin, i.e., 
\begin{eqnarray}
    \lim_{u\rightarrow0}\sigma(u)&=&\sigma_0,\\
    \lim_{u\rightarrow\pm\infty}\sigma(u)&=&0,
\end{eqnarray}
where $\sigma_0$ is a constant, which accounts for the maximum value of the surface tension. These conditions guarantee a stress tensor concentrated around the catenoid throat. Indeed, since the surface is asymptotic flat, the curvature and strain effects should vanish as $u\rightarrow\infty$.

\begin{figure*} 
       \begin{minipage}[b]{0.48 \linewidth}
           \includegraphics[width=\linewidth]{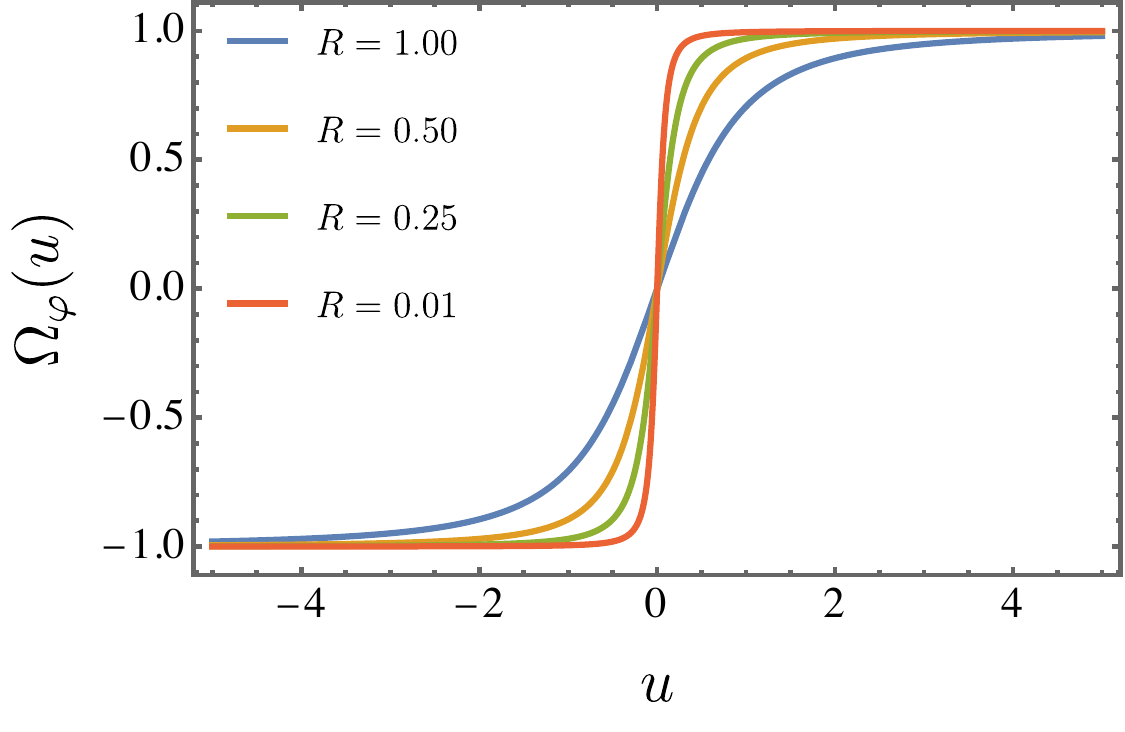}\\
           \caption{Geometric connection $\Omega_\phi$ for $R=0.1$ and $R=1$.}
          \label{geometricconnection}
       \end{minipage}\hfill
       \begin{minipage}[b]{0.515 \linewidth}
           \includegraphics[width=\linewidth]{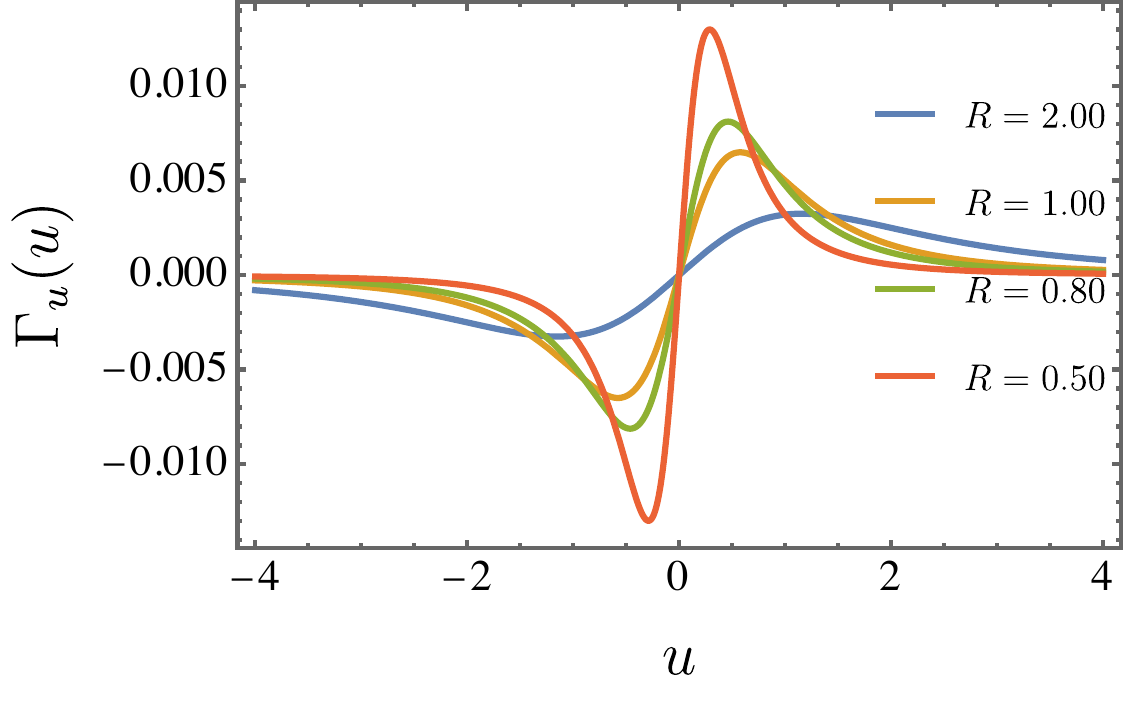}\\
           \caption{Strain vector $\Gamma_u$ for $R=0.1$ and $R=1$.}
           \label{strainvector}
       \end{minipage}
\end{figure*}

We assume that the mechanical properties of the surface are in the linear elastic regime. Thereby, the stress and the strain tensors are related by \cite{vozmediano}
\begin{equation}
    \sigma_{ij}=\lambda \theta g_{ij} + 2\mu u_{ij},
\end{equation}
where $\lambda$ and $\mu$ are the Lam\'{e} coefficients and $\theta=u^{i}_{i}$ is the trace of the strain tensor. From the ansatz employed in eq.(\ref{strainansatz}), we obtain the strain tensor as
\begin{equation}
\label{straintensor}
    u^{i}_{j}=\frac{1}{2(\lambda+\mu)}\sigma(u)\delta^{i}_{j}.
\end{equation}
The form of the strain tensor in eq.(\ref{straintensor}) shows that the deformations undergone by the surface are isotropic and concentrated around the catenoid throat. The position--dependent Fermi velocity tensor $v_{ij}$ can be written as 
\begin{equation}
\label{velocitytensor}
v_i^j=v(u)\delta_i^j,    
\end{equation}
where the position--dependent Fermi velocity function
$v(u)$ is given by
\begin{equation}
\label{dependentvelocity}
    v(u)=v_0\left[1-\frac{\beta}{2}\frac{\sigma(u)}{(\lambda+\mu)}\right].
\end{equation}
Accordingly, the components of the pseudo--vector potential $\Gamma_i$ are
\begin{eqnarray}
    \Gamma_u&=&-\frac{\beta}{4}\frac{\sigma'(u)}{(\lambda+\mu)},\\
    \Gamma_{\phi}&=&0.
\end{eqnarray}
In this work, we assume the isotropic stress function $\sigma(u)$ as
\begin{equation}
\label{stressfunction}
    \sigma(u)=\sigma_0 \frac{R^2}{R^2 + u^2}.
\end{equation}
We see that $\sigma(u)$ becomes even more concentrated when $R\rightarrow 0$, as it is shown in fig. (\ref{stress}). On the other hand, in fig. (\ref{velocity}), we show the plot of the Fermi velocity function $v(u)$ for different values of $\beta$. Remarkably, it decreases for the regions close to the wormhole throat (high curvature). Such a feature was already found in a smooth ripple curved graphene layer \cite{contijo}.

In addition, the behavior of the geometric spinor connection $\Omega_\varphi$ and the strain vector $\Gamma_u$ are shown in fig.(\ref{geometricconnection}) and fig.(\ref{strainvector}), respectively. Note that both terms are parity odd functions with respect to the $u$ coordinate. In this sense, both potentials yield to barrier between the lower $u<0$ and upper $u>0$ layers. However, despite this similarity, the strain potential given by eq.(\ref{strainvectorfield}) and the spinor potential given by eq.(\ref{spinorconnection}) have different natures, forms and components. Therefore, these potentials produce different effects on the Dirac electron, as we shall see in the next section.  



\subsection{Magnetic vector potential}

\begin{figure*} 
       \begin{minipage}[b]{0.48 \linewidth}
           \includegraphics[width=\linewidth]{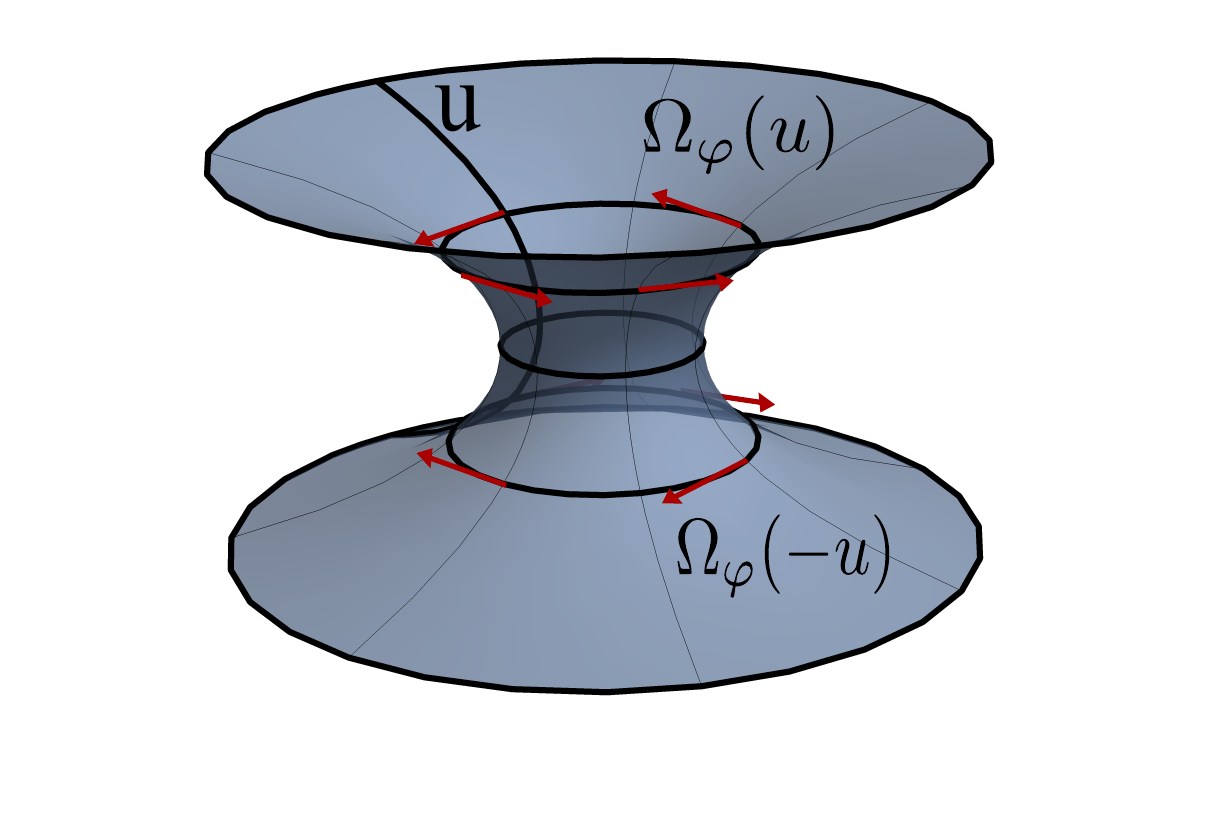}\\
           \caption{Spin connection $\Omega_\phi$ for $R=0.1$ and $R=1$. 
           This curvature potential exhibit a parity--odd angular configuration.}
          \label{spinconnectionfigure}
       \end{minipage}\hfill
       \begin{minipage}[b]{0.48 \linewidth}
           \includegraphics[width=\linewidth]{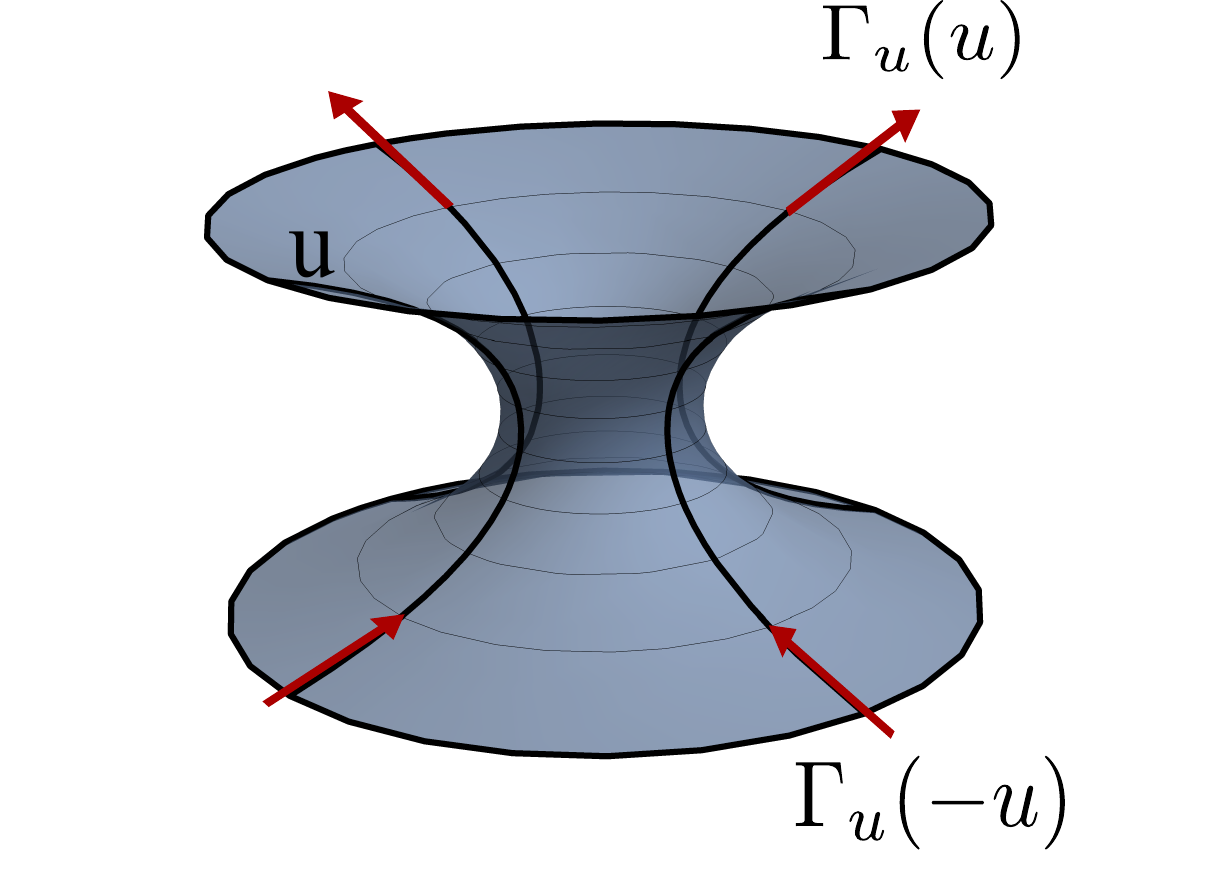}\\
           \caption{Strain vector $\Gamma_u$ for $R=0.1$ and $R=1$. The strain--driven potential has a components only along the meridian.}
           \label{strainvectorfigure}
       \end{minipage}
\end{figure*}

\begin{figure*} 
       \begin{minipage}[b]{0.48 \linewidth}
           \includegraphics[width=\linewidth]{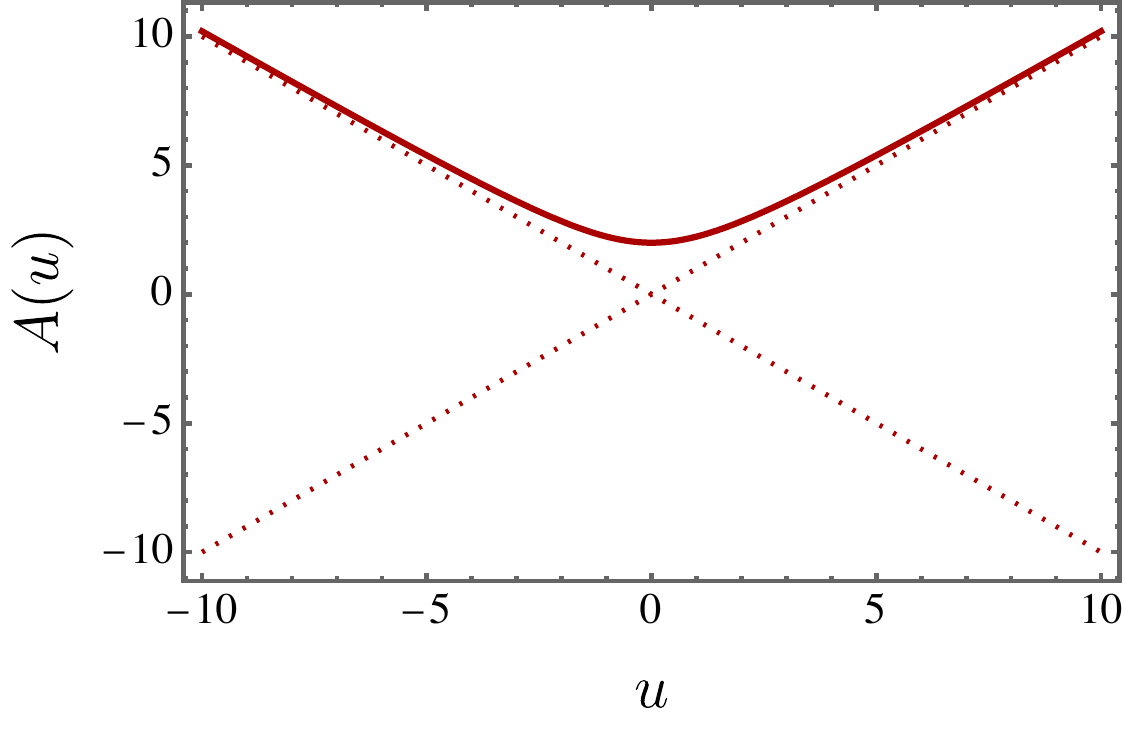}\\
           \caption{Vector potential angular component for an uniform magnetic field.}
          \label{vectorpotential}
       \end{minipage}\hfill
       \begin{minipage}[b]{0.515 \linewidth}
           \includegraphics[width=\linewidth]{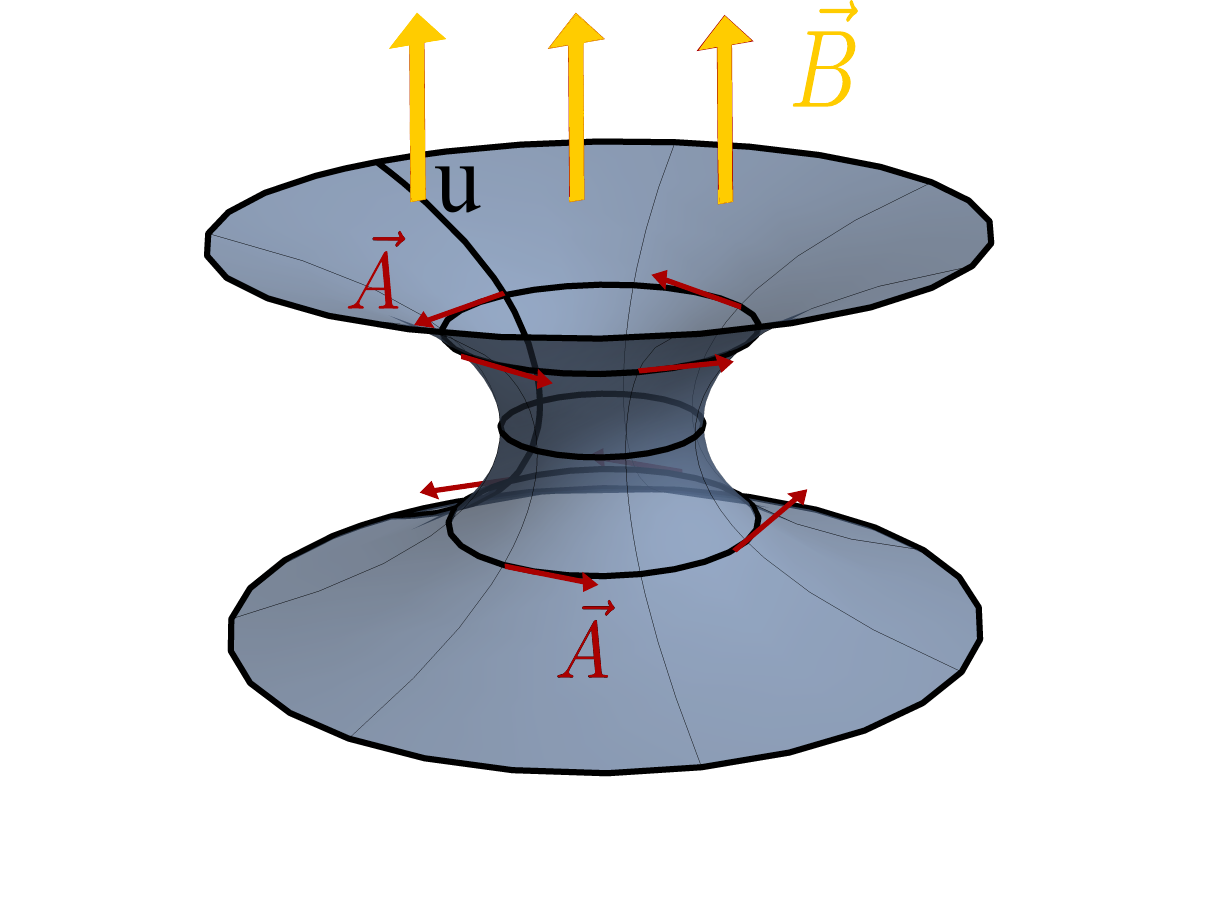}\\
           \caption{Magnetic vector potential $\vec{A}$ on the surface. }
           \label{vectorpotentialfigure}
       \end{minipage}
\end{figure*}

Now, let us see how the external magnetic field $\vec{B}$ modifies the Hamiltonian. 
For a uniform magnetic field
$\vec{B}=B\hat{k}$, the vector potential $\vec{A}$ is given by $\vec{A}=\frac{1}{2}\vec{B}\times \vec{r}$. Using the coordinates system in eq.(\ref{coordinates}), we obtain
\begin{equation}
\label{Avector}
    \vec{A}=\frac{B}{2}\sqrt{R^2 + u^2}\hat{e}_2 . 
\end{equation}
For $R=0$, the expression for the vector potential in eq.(\ref{Avector}) reduces to $\vec{A}=\pm \frac{B}{2}u\hat{e}_2$, as found in Ref.(\cite{diracmagnetic1}).
For $u\gg R$, it yields  $A^{2}\approx \frac{B}{2}u$, as in a conical surface \cite{diracmagnetic1} and on the flat plane \cite{diracplanar}. In addition, at $u=0$, it has a finite value $A^{2}=\frac{BR}{2}$.
Since $\vec{e}_\phi =\sqrt{R^2 +u^2}\hat{e}_2$, the component of $\vec{A}$ in coordinates is given by $A^\phi =\frac{B}{2}$. Accordingly, for the covariant component $A_\phi = g_{\phi\phi}A^{\phi}$, we have
\begin{equation}
    A_\phi = \frac{B}{2}(R^2 +u^2).
\end{equation}
A similar expression for the vector potential was found for the discontinuous graphene wormhole \cite{wormhole3}, except for the presence of the throat radius $R$.
The electromagnetic potential displayed in fig.(\ref{vectorpotential}) is parity even, in contrast with the geometric spinor connection shown in fig.(\ref{spinconnectionfigure}). Also, the strain vector in fig.(\ref{strainvectorfigure}) turns out to have a parity odd configuration.

\section{Effective Hamiltonian}
\label{section4}

Once we have discussed all those interactions acting upon the electron, i.e., the curved geometry, strain and external magnetic field, let us now obtain the respective Hamiltonian. By
collecting all the interactions, the effective Hamiltonian becomes
\begin{eqnarray}
\label{effecham}
    \nonumber\mathcal{H}_D&=&-i\hbar v_0\left\{\sigma_1 \left[v(u)\partial_u+\bar{\beta}\sigma' +\frac{u}{2(R^2+u^2)}\right]\right.\\
    &&\left.-\frac{i\sigma_2}{\sqrt{R^2+u^2}}[v(u)\partial_{\phi}+eA_\phi]\right\},
\end{eqnarray}
where $\bar{\beta}=\frac{\beta}{4(\lambda+\mu)}$. Notice that the strain vector and the spinor connection modifies the Dirac equation leading to a canonical momentum along the $u$ direction of form
\begin{equation}
    \hat{P}_u = - i\hbar v_0\left[v(u)\partial_u+\bar{\beta}\sigma' +\frac{u}{2(R^2+u^2)}\right].
\end{equation}
Additionally, along the angular $\phi$ direction, the canonically conjugate momentum is modified by
\begin{equation}
    \hat{P}_\phi = -i\hbar v_0\frac{1}{\sqrt{R^2+u^2}}[v(u)\partial_{\phi}].
\end{equation}
In this manner, the effective Hamiltonian can be rewritten in the familiar form $\mathcal{H}_D = v_0 \vec{\sigma}\cdot (\vec{P}-e\vec{A}),
$ where $\vec{P}=(P_u , P_\phi)$ are the canonically conjugate momenta and $\vec{\sigma}=(\sigma_1 , \sigma_2)$ are the flat Pauli matrices.

The expression in eq.(\ref{effecham}) depends only on the coordinate $u$. The symmetry of the Hamiltonian with respect to the angular $\phi$ variable is the result of the surface axial symmetry. Thus, the wave function should also inherit this symmetry. In fact, consider the angular momentum operator with respect to the $\mathrm{z}$ axis, $\hat{L}_z = -i\hbar \frac{\partial}{\partial\phi}$ such that, $\hat{L}_z \psi = \hbar l \psi$, where $l$ is the orbital angular momentum with respect to the $\mathrm{z}$ axis. For a non--relativistic and spinless electron on the graphene wormhole, an axisymmetric wave function can be written as $\psi(u,\phi)=e^{i l\phi}\psi(u)$ \cite{euclides}. However, as it is well--known for the relativistic electron, $\hat{L}_z$ no longer commute with $\mathcal{H}_D$, although the total angular momentum operator along the $z$  direction $\hat{J}_z =\hat{L}_z + \hat{S}_z$ does \cite{diracplanar}. Since $\hat{S}_3 = \frac{\hbar}{4}[\gamma^{1},\gamma^{2}]$, then the spin operator with respect to the $\mathrm{z}$ axis is given by
\begin{equation}
S_z =-i\frac{\hbar}{2}\sigma_3. \end{equation}

Here, the total angular momentum operator has the form $\hat{J}_z=-i\hbar \left(\frac{\partial}{\partial\phi} + \frac{1}{2}\sigma_3 \right)$, where $\hat{J}\psi = m\hbar \psi$ and $m=l\pm 1/2$ \cite{diracplanar2}.
Therefore, considering the axial symmetry on the spinorial wave function, so that \cite{wormhole4,diracplanar2}
\begin{equation}
    \Psi(u,\phi)=e^{im\phi}\psi(u),
\end{equation}
the Dirac equation $\mathcal{H}_D \Psi=E\Psi$ leads to the Dirac equation $\tilde{\mathcal{H}}_D \psi=E\psi$, in which the effective Hamiltonian simplifies to
\begin{widetext}
\begin{eqnarray}
\label{effectivehamiltonian2}
    \tilde{\mathcal{H}}_D&=&-i\hbar v_0\left(\begin{array}{cc}
        0 & v(u)\partial_u+\bar{\beta}\sigma'+\frac{u}{2(R^2+u^2)}-\frac{[v(u)m+eA_\phi]}{\sqrt{R^2+u^2}} \\
        v(u)\partial_u+\bar{\beta}\sigma' +\frac{u}{2(R^2+u^2)}+\frac{[v(u)m+eA_\phi]}{\sqrt{R^2+u^2}}  & 0
    \end{array}\right).
\end{eqnarray}
\end{widetext}
The effective Hamiltonian in eq.(\ref{effectivehamiltonian2}) has parity-even and parity-odd potentials.

\begin{figure*}[htb] 
       \begin{minipage}[b]{0.49 \linewidth}
           \includegraphics[width=\linewidth]{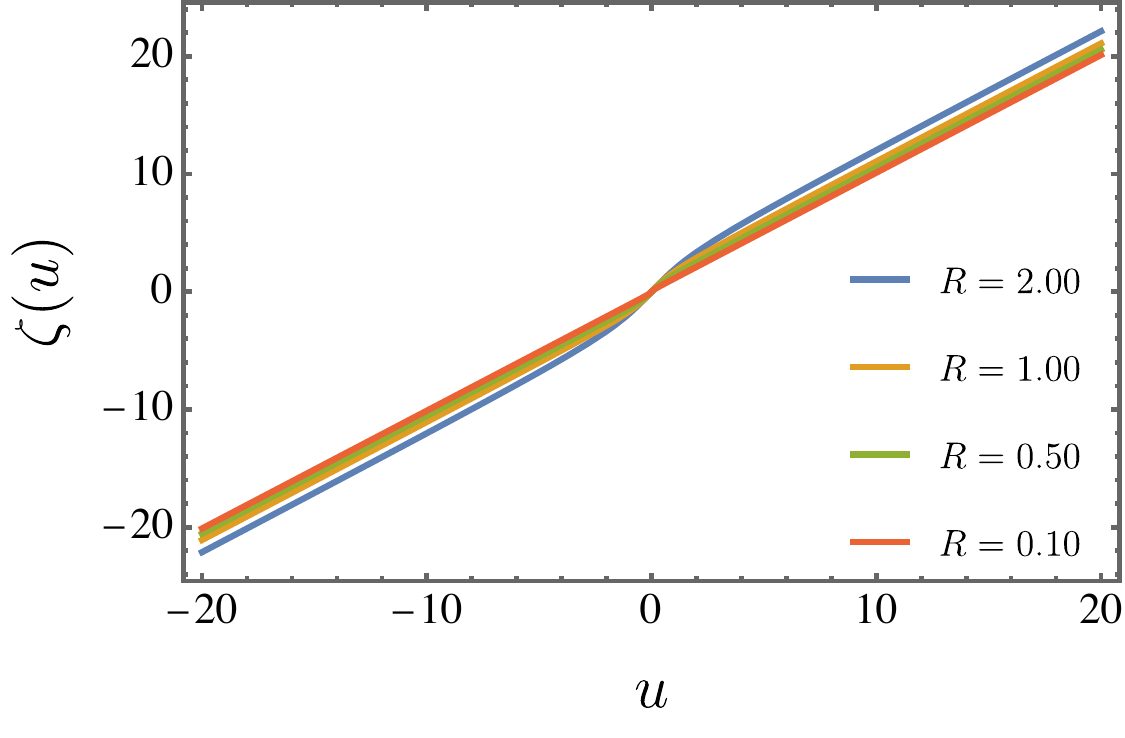}\\
           \caption{Change of coordinate $\zeta=\zeta(u)$ for $\beta=1$.}
          \label{changeofcoordinate}
       \end{minipage}\hfill
       \begin{minipage}[b]{0.48 \linewidth}
           \includegraphics[width=\linewidth]{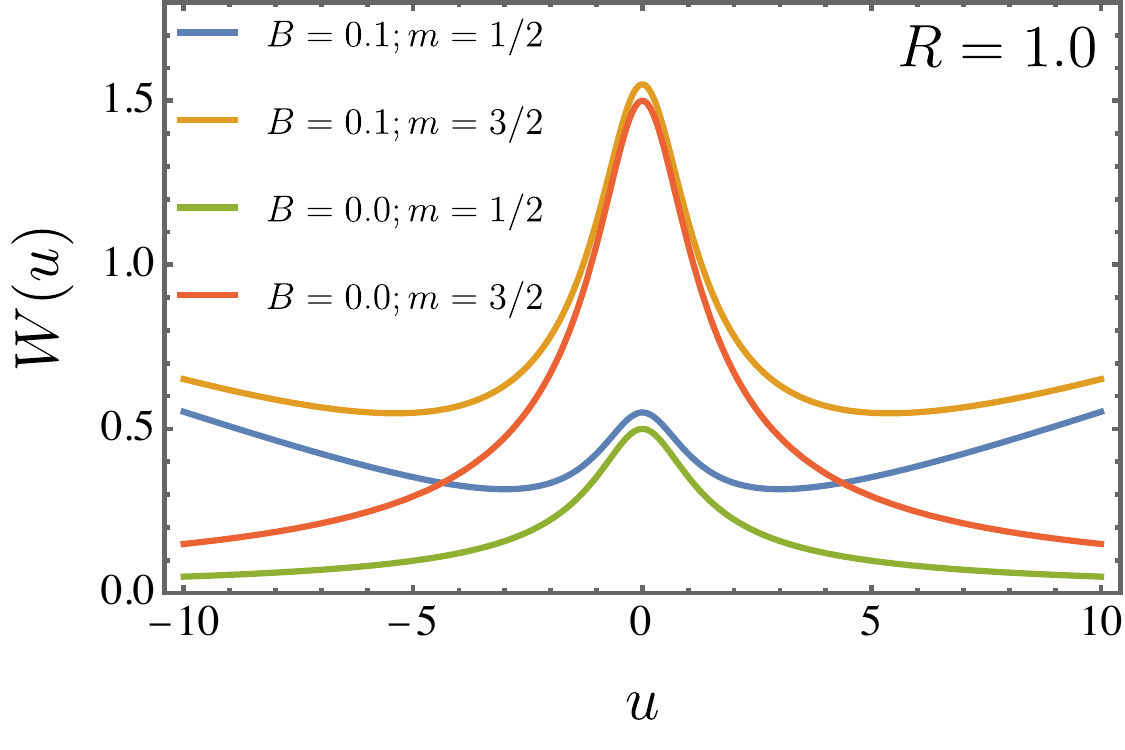}\\
           \caption{Superpotential $W(u)$ for $R=1$, $B=0$.}
           \label{superpotentialb0}
       \end{minipage}
\end{figure*}

In eq.(\ref{effectivehamiltonian2}), the position--dependent velocity function $v(u)$ multiplies the partial derivative $\partial_u$.
By writing $v\frac{d}{du}=\frac{d}{d\zeta}$, for $v(u)$ given by eq.(\ref{dependentvelocity}), we have $\zeta=u-\frac{\bar{\beta}R}{\sqrt{2(\bar{\beta}-2)}} \,\tanh^{-1}\left(\frac{\sqrt{2}u}{R\sqrt{(\bar{\beta}-2)}}\right)$. Unfortunately, this relation can not be inverted analytically, as one seeks to rewrite eq. (\ref{effectivehamiltonian2}) in terms of variable $\zeta$. However, a graphic analysis reveals that, for $0\leq \bar{\beta}\leq 1$, there is only a small difference between $\zeta$ and $u$, as shown in fig. (\ref{changeofcoordinate}) for $\bar{\beta}=1$. Indeed, taking the limit $\bar{\beta}\rightarrow 1$, the function $\zeta$ becomes $\zeta=u+\frac{R}{\sqrt{2}}\tan^{-1}{\sqrt{2}u}$.
Note that for $u\rightarrow\pm\infty$, $\zeta - u \rightarrow \pm\frac{R}{\sqrt{2}}$, and thus $\frac{\zeta - u}{u}\rightarrow 0$. Therefore, near the origin $\zeta\approx u$ and asymptotically $\zeta$ is only $u$ shifted by a constant amount, as we can also see by the graphic in fig.(\ref{changeofcoordinate}). For $1<\bar{\beta}<2$, the difference $\zeta-u$ increases, though it is still finite and $\frac{(\zeta-u)}{u}\rightarrow 0$, as $u\rightarrow\infty$. At $\bar{\beta}=2$ the function $\zeta$ is not defined and for $\bar{\beta}>2$ the function $\zeta$ is only defined in the interval $|u|<\frac{R}{\sqrt{2}}\sqrt{\bar{\beta}-2}$. Since we are interested in solutions for all $-\infty<u<\infty$ and we are considering small strain effects, i.e., $\bar{\beta}\sigma_0 \leq 1/2$, then we restrict ourselves to the interval $0\leq \bar{\beta}\leq 1$. Accordingly, since the position-dependence of the Fermi velocity is small for $0\leq \bar{\beta}\leq 1$, for the sake of simplicity we adopt $\zeta\approx u$ from now on.

As a result, the Hamiltonian simplifies to
\begin{widetext}
\begin{eqnarray}
\label{effectivehamiltonian3}
    \tilde{\mathcal{H}}_D&=&-i\hbar v_0\left(\begin{array}{cc}
        0 & \partial_u+\bar{\beta}\sigma'+\frac{u}{2(R^2+u^2)}-\frac{[m+eA_\phi]}{\sqrt{R^2+u^2}} \\
        \partial_u+\bar{\beta}\sigma' +\frac{u}{2(R^2+u^2)}+\frac{[m+eA_\phi]}{\sqrt{R^2+u^2}}  & 0
    \end{array}\right).
\end{eqnarray}
\end{widetext}
The effective Hamiltonian in eq.(\ref{effectivehamiltonian3}) shows clearly the distinctive interaction terms arising from the strain $\bar{\beta}\sigma'$, from the geometric connection $\frac{u}{2(R^2+u^2)}$, centrifugal term $\frac{m}{\sqrt{R^2+u^2}}$
and the electromagnetic coupling $\frac{eA_\phi}{\sqrt{R^2+u^2}}$. In the next section, we explore the effects of each interaction.


\section{Supersymmetric analysis}
\label{section5}

In this section, we employ a supersymmetric quantum mechanical approach \cite{ozlem,ozlem2} to explore the features of the effective Hamiltonian in eq.(\ref{effectivehamiltonian3}) and find the solutions of the Dirac equation.

From the effective eq.(\ref{effectivehamiltonian3}), it leads to
\begin{equation}
\label{diracequation}
   \tilde{\mathcal{H}}_D\psi=\epsilon\psi,   
\end{equation}
where the spinor $\psi=\left(\begin{array}{cc}
         \psi_1 \\
         \psi_2, 
\end{array}\right)$. The effective Dirac equation (\ref{diracequation}) can be written as
\begin{eqnarray}
\label{diraczeroelectric}
    \left(\begin{array}{cc}
        0 & i\mathcal{O}_2 \\
       i\mathcal{O}_1 & 0  
    \end{array}\right)\left(\begin{array}{cc}
         \psi_1 \\
         \psi_2, 
    \end{array}\right) = \epsilon \left(\begin{array}{cc}
         \psi_1 \\
         \psi_2, 
    \end{array}\right),
\end{eqnarray}
where the first-order operators $\mathcal{O}_{1,2}$ are defined as
\begin{equation}
\mathcal{O}_{1,2}=\frac{\mathrm{d}}{\mathrm{d}u}+\bar{\beta}\sigma' + \frac{u}{2(R^2 + u^2)}\pm \frac{(m+eA_{\phi})}{\sqrt{R^2+u^2}}.
\end{equation}
By performing the change on the wave function of the form
\begin{equation}
\label{wavefunctionchange}
    \psi_{1,2}(u)=(R^2+u^2)^{-1/4}e^{-\bar{\beta}\sigma(u)}\chi_{1,2}(u),
\end{equation}
the Dirac equation yields a decoupled equations for the $\chi_1$ and $\chi_2$ in a Klein--Gordon like form
\begin{eqnarray}
\label{kleingordonlikeequation}
    -\chi_{1,2}'' + U_{eff1,2}^2\chi_{1,2} = \epsilon^2 \chi_{1,2},
\end{eqnarray}
where $\epsilon=\frac{E}{\hbar v_0}$ is the electron momentum and the squared effective potential is given by
\begin{equation}
\label{effectivepotential}
    U_{eff 1,2}^2 = \left(\frac{(m+eA_\phi)}{\sqrt{R^2 +u^2}}\right)^2\mp\left(\frac{(m+eA_\phi)}{\sqrt{R^2 +u^2}}\right)'.
\end{equation}

The Klein--Gordon--like expression present in eq. (\ref{kleingordonlikeequation}) has the structure of a so--called supersymmetric quantum mechanics, whose superpotential $W$ is given by
\begin{eqnarray}
\label{superpotential}
W=\frac{m+eA_{\phi}}{\sqrt{R^2+u^2}}.
\end{eqnarray}
Note that eq.(\ref{superpotential}) is given by the spin--curvature potential $\frac{m}{\sqrt{R^2+u^2}}$ and the magnetic coupling term $\frac{eA_{\phi}}{\sqrt{R^2+u^2}}$ present in the Dirac equation.

The SUSY--like form of the squared effective potential 
\begin{equation}
U_{eff 1,2}^{2}=W^2 \mp W'    
\end{equation}
enables us to rewrite the decoupled system of second--order Klein--Gordon like eq.(\ref{kleingordonlikeequation}) as
\begin{eqnarray}
\label{susyeq}
    a^{\dagger}a \chi_1 = \epsilon^2 \chi_1 \\
    a a^{\dagger} \chi_2 = \epsilon^2 \chi_2 ,
\end{eqnarray}
where $a=\frac{d}{du} + W$ and $a^{\dagger}=-\frac{d}{du}+W$
are first--order differential operators \cite{ozlem2}. The so-called superpartner squared Hamiltonians $H_{1}^{2} =a^{\dagger}a$ and $H_{2}^{2} =aa^{\dagger}$ satisfy $H_{2}^{2}=(H_{1}^{2})^{\dagger}$ \cite{ozlem2}.

A remarkable feature of a quantum mechanical SUSY--like eq.(\ref{susyeq}) is the existence of a nonvanishing ground state for $\epsilon=0$, known as the zero mode \cite{wormhole,picak}. Indeed, for $\epsilon=0$, the conditions $a\chi_1 =0$ and $a^{\dagger}\chi_2$ yield to
\begin{equation}
    \psi^{0}_{1,2}(u)=(R^2+u^2)^{-1/4}e^{-\bar{\beta}\sigma(u)}e^{\mp\int{W(u')du'}}.
\end{equation}
Since the superpotential $W$ is related to the spin--curvature coupling and the external magnetic field, the zero mode is related to the flux of curvature and magnetic field near the throat \cite{wormhole}.

The factor $(R^2 + u^2)^{-1/4}$ is due to the normalization condition on the surface. Indeed, the normalization condition takes the form
\begin{equation}
    1=\int_{-b}^{b}\int_{0}^{2\pi}{||\psi||^2 (R^2 + u^2)^{1/2}\mathrm{d}u\mathrm{d}\phi},
\end{equation}
where $-b<u<b$.
Despite both the curved geometry and the strain reduces the wave function amplitude for $u\rightarrow \pm \infty$, the curvature damps the amplitude by a power-law factor whereas the strain damps it by an exponential factor.

Another noteworthy feature of the Dirac equation in curved surface is related to the geometric phase \cite{furtado}. Indeed, the holonomy operator $U(\phi)=e^{\int_{0}^{\phi}{\Omega_i dx^{i}}}$,
where $\Omega_i$ is the spin connection in eq.(\ref{spinorconnection}) leads to
\begin{equation}
\label{holonomyoperator}
    U(\phi)=e^{-\frac{i}{2}\frac{u}{R^2 + u^2}\sigma_3 \phi}.
\end{equation}
This geometric phase reflects the change on the wave function when the fermion performs a $2\pi$ rotation for a given $u$ \cite{mobius2}. It is a kind of geometric Aharonov--Bohm effect driven by the curvature instead of the magnetic field \cite{furtado}.
By applying the geometric phase operator $U(\phi)$, i.e., $\psi' =U(\phi)\psi$, the Dirac equation $\tilde{\mathcal{H}}_D \psi = E \psi$ simplifies to $\hat{\mathcal{H}}_D \psi' = E \psi'$, where 
the simplified Hamiltonian $\hat{\mathcal{H}}_D$ is given by
\begin{eqnarray}
\label{effectivehamiltonian4}
    \tilde{\mathcal{H}}_D&=&-i\hbar v_0\left(\begin{array}{cc}
        0 & \partial_u+\bar{\beta}\sigma'-\frac{[m+eA_\phi]}{\sqrt{R^2+u^2}} \\
        \partial_u+\bar{\beta}\sigma' +\frac{[m+eA_\phi]}{\sqrt{R^2+u^2}}  & 0
    \end{array}\right).
\end{eqnarray}
Thus, the curvature effects of the curved surface can be encoded into the geometric phase in the operator $U(\phi)$ given by eq.(\ref{holonomyoperator}).


It is important to highlight that the effects of the strain, curved geometry and external magnetic fields are rather distinct. The strain and curved geometry provide the geometric phase, whereas the centrifugal and the magnetic field yield the SUSY--like symmetry. In the following, we explore the effects of each term on the electronic states.


\begin{figure}
    \centering
    \includegraphics[scale=0.4]{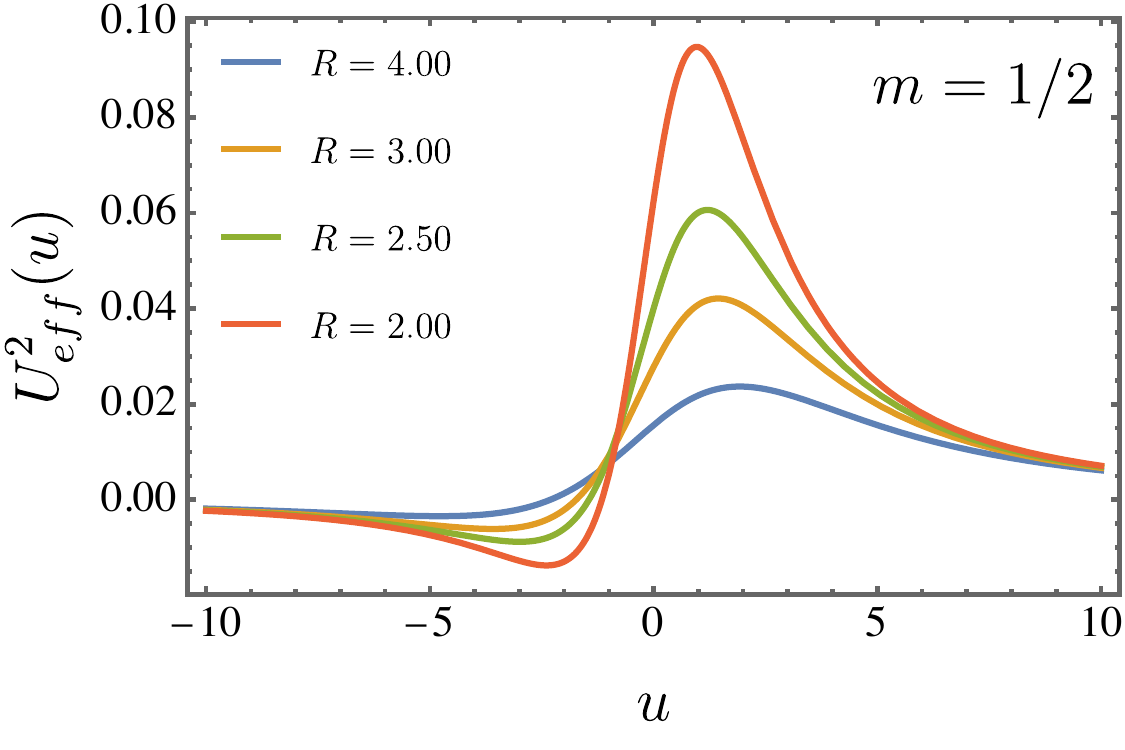}
     \includegraphics[scale=0.4]{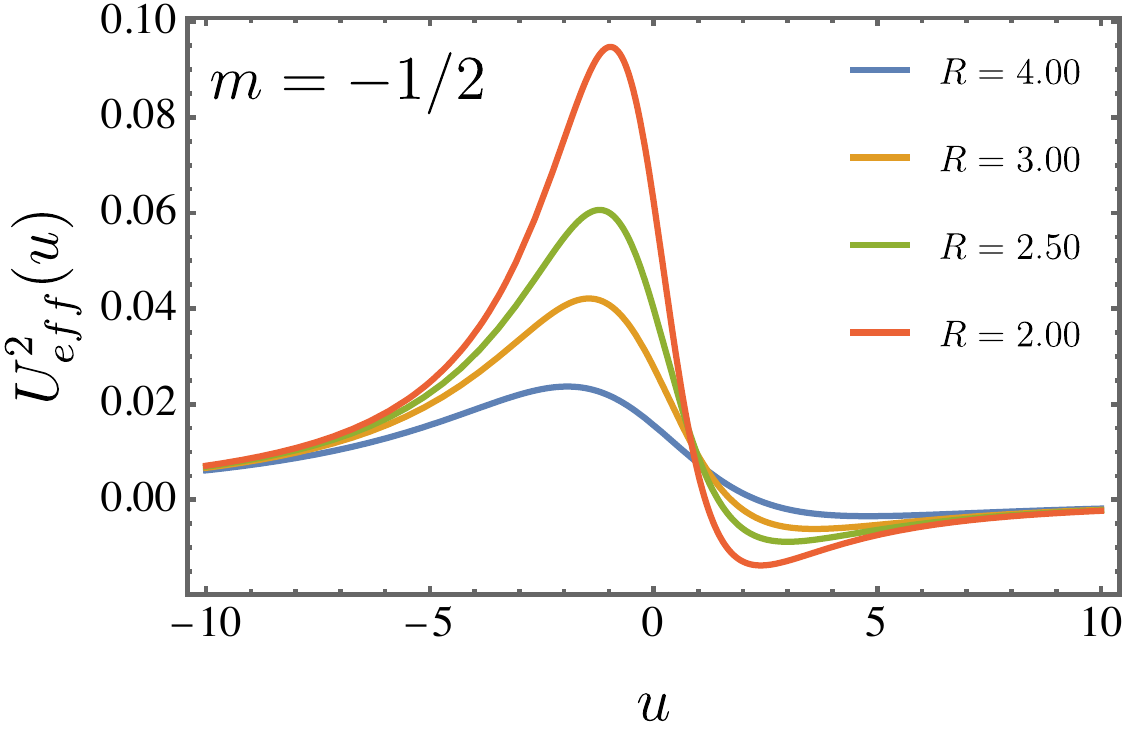}
    \caption{Squared effective potential for $\bar{\beta}=1$, $m=1/2$ (on the left panel), $m=-1/2$ (on the right panel) for $B=0$.}
    \label{effectivepotentialb0}
\end{figure}

\subsection{No external magnetic field}

In the absence of a magnetic field, i.e., for $B=0$, the superpotential has a simple form 
\begin{equation}
 W(u)=\frac{m}{\sqrt{R^2 + u^2}},   
\end{equation}
whose behavior is plotted in fig.\ref{superpotentialb0}. The symmetric centrifugal barrier of the superpotential leads to an asymmetric potential for $U^{2}_{eff}$, as shown in fig.(\ref{effectivepotentialb0}). Note the dependence of the squared potential on the total angular momentum $m$. This one is similar to that one encountered in the context of a Dirac electron constrained to a helicoid potential \cite{watanabe}.

\begin{figure*} 
       \begin{minipage}[b]{0.48 \linewidth}
           \includegraphics[width=\linewidth]{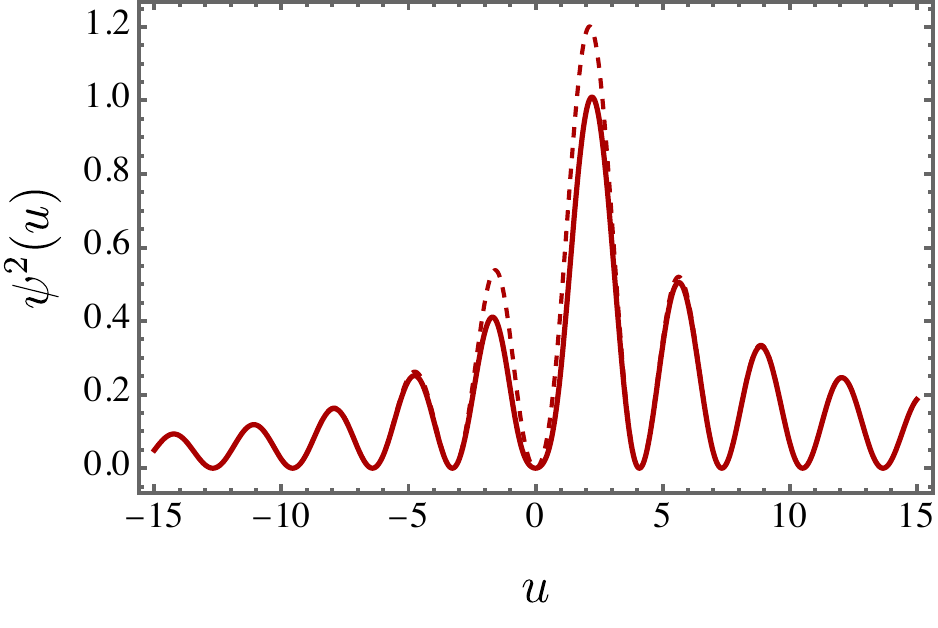}\\
           \caption{Density of states for $\epsilon=1$ and $m=1/2$.}
          \label{dfmone}
       \end{minipage}\hfill
       \begin{minipage}[b]{0.48 \linewidth}
           \includegraphics[width=\linewidth]{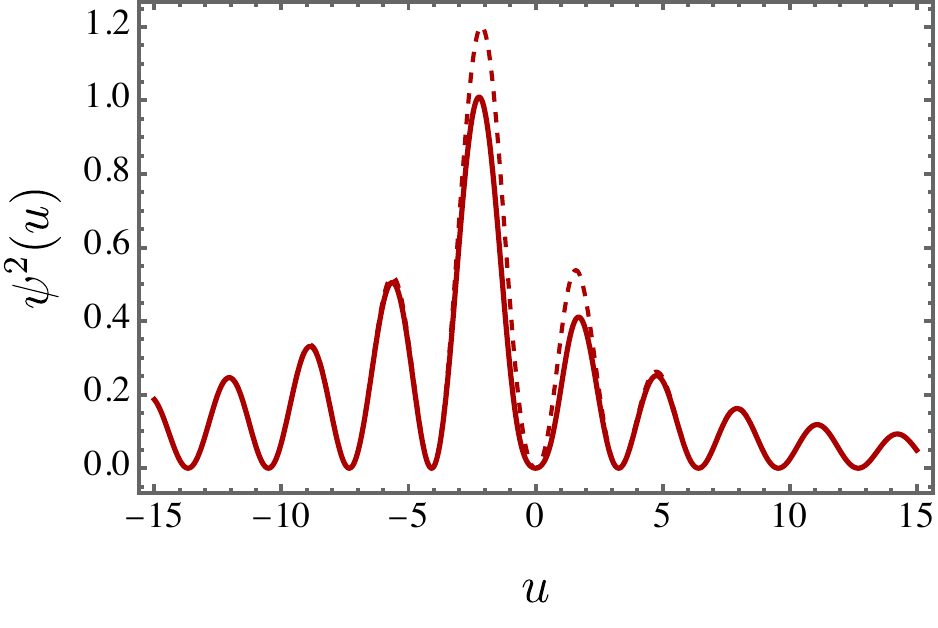}\\
           \caption{Density of states for $\epsilon=1$ and $m=-1/2$.}
           \label{dfmminusone}
       \end{minipage}
\end{figure*}

For $m\neq 0$, the Klein--Gordon like eq.(\ref{kleingordonlikeequation}) reads
\begin{equation}
\label{mneq0}
    -\chi_{1,2}'' + \left(\frac{m^2}{R^2 + u^2} \pm m\frac{u}{(R^2 + u^2)^{3/2}}\right)\chi_{1,2} = \epsilon^2 \chi_{1,2},
\end{equation}
It is worthwhile to mention that the effective potential 
\begin{equation}
U^{2}_{eff 1,2}=\frac{m^2} {R^2 + u^2} \pm m\frac{u}{(R^2 + u^2)^{3/2}},    
\end{equation}
couples the angular momentum quantum number $m$ and the curved geometry terms.
The second term in the potential $\frac{m}{(R^2 + u^2)^{3/2}}$ breaks the symmetry $m\rightarrow -m$. Indeed, we can obtain the $\chi_2$ spinor component from $\chi_1$ by performing the change $m\rightarrow -m$.

For $u\gg R$, the potential tends to $U_{eff 1,2}\approx\frac{m(m\pm 1)} {u^2}$. Accordingly, the eq.(\ref{mneq0}) has the asymptotic solution
\begin{equation}
\label{masymptotic}
    \chi_{1,2} \approx \sqrt{u}\left(c_1  J_{\frac{2m\pm 1}{2}}(\epsilon u) + c_2 Y_{\frac{2m\pm 1}{2}}(\epsilon u)\right),
\end{equation}
where $J_n (x)$ and $Y_n (x)$ are the Bessel function of the first kind and the second kind, respectively. In this region, the solution in eq.(\ref{masymptotic}) resembles the one found for the Dirac equation in other graphene wormhole geometries outside the throat \cite{wormhole4}.

Note the presence of the total angular momentum number $m=l+\frac{1}{2}$ in the order of the Bessel functions.
For $u\rightarrow\pm\infty$, the potential vanishes and thus, the $\chi_1$ function tends to $A\sin(\epsilon u)$, as for the $m=0$ states. Therefore, the interaction between angular momentum and curvature is concentrated around the graphene wormhole throat.

For $R=1$ and $\epsilon=1$, we numerically solved the eq.(\ref{mneq0}) and the resulting squared wave function was plotted in fig.(\ref{dfmone}) for $m=1$ and in the fig.(\ref{dfmminusone}) for $m=-1$. By changing $m\rightarrow -m$, the density of state is shift from the upper $(u>0)$ into the lower sheet $(u<0)$. Moreover, for $\beta=1$ (solid line) the amplitude is smaller than for $\beta=0.5$ (dashed line).


The zero mode, i.e., for $\epsilon=0$ is given by
\begin{equation}
    \psi^{0}_{1,2}(u)=(R^2+u^2)^{-1/4}e^{-\bar{\beta}\sigma(u)}e^{\mp m\sinh^{-1}(u/R)},
\end{equation}
where is evident the chiral symmetry breaking $m\rightarrow -m$ and the parity-odd behaviour of this ground state. Indeed, as shown in the fig. the states for $m>0$ are suppressed in the upper layer whereas the $m<0$ states are suppressed in the lower layer. A similar chiral separation was also found for the massless Dirac field in a helicoidal graphene strip \cite{atanasov}.
\begin{figure*} 
       \begin{minipage}[b]{0.48 \linewidth}
           \includegraphics[width=\linewidth]{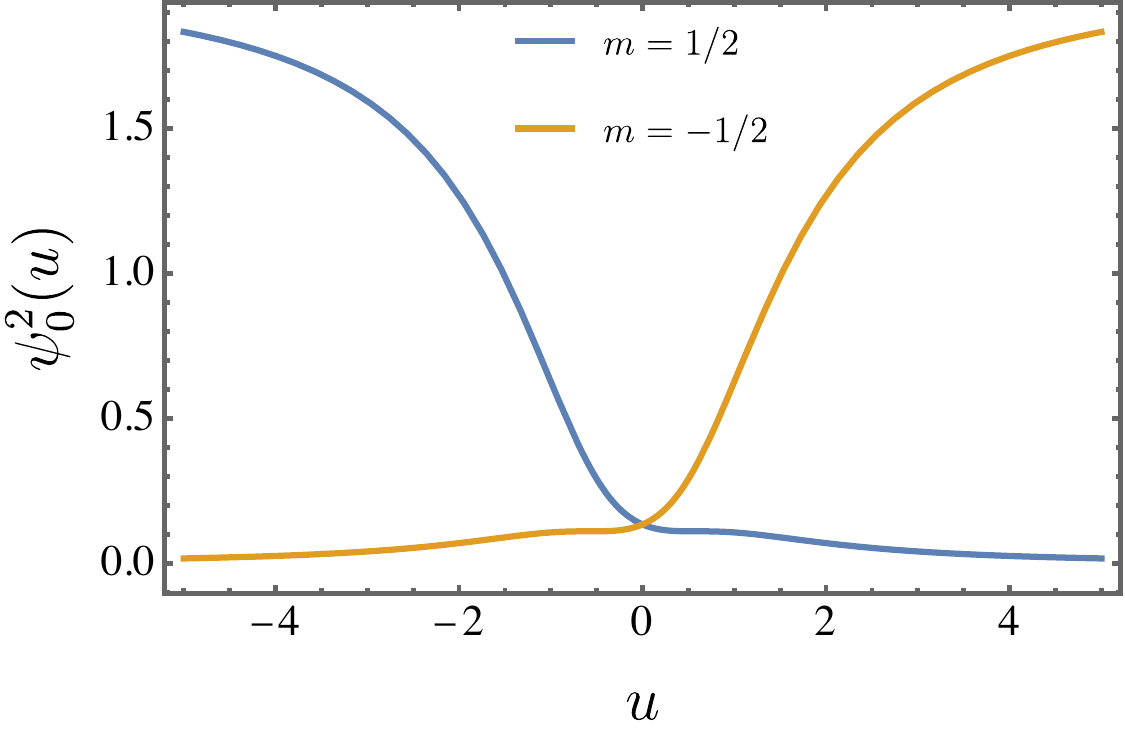}\\
           \caption{Zero mode for $m=\pm 1/2$.}
          \label{zeromode12}
       \end{minipage}\hfill
       \begin{minipage}[b]{0.48 \linewidth}
           \includegraphics[width=\linewidth]{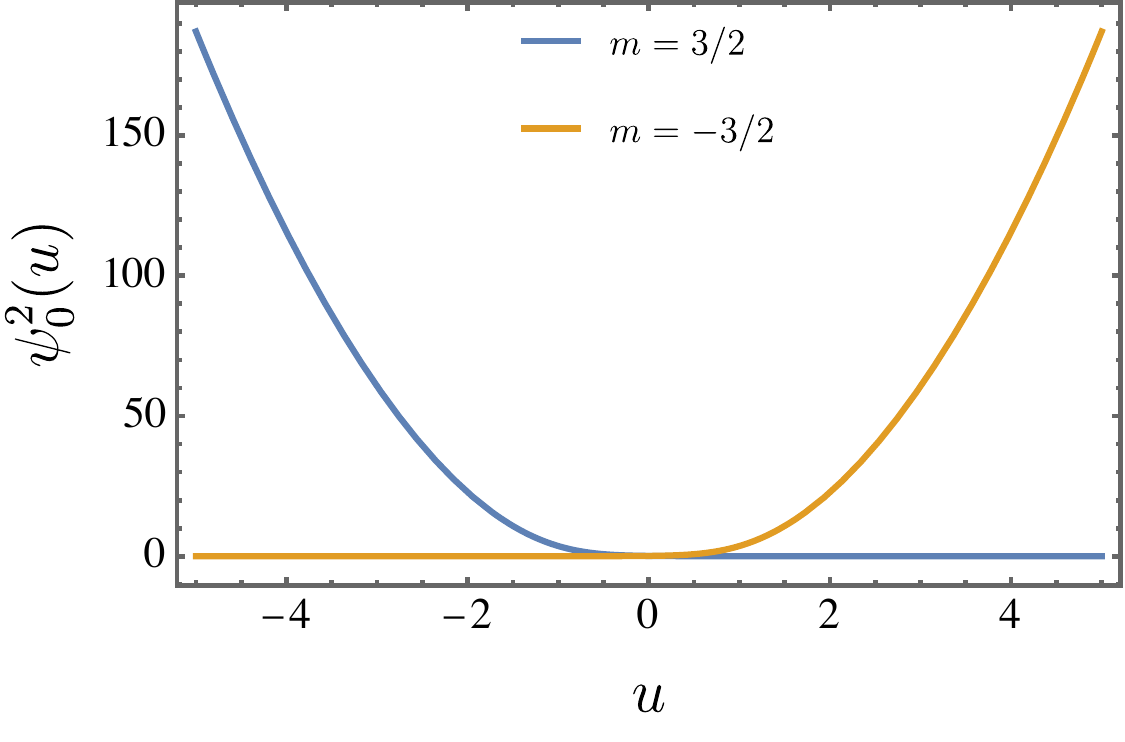}\\
           \caption{Zero mode $m=\pm 3/2$.}
           \label{zeromode32}
       \end{minipage}
\end{figure*}

\subsection{Constant magnetic field}

Now let us consider the additional effects from the uniform magnetic field. The respective Klein-Gordon like equation becomes
\begin{equation}
\label{kleingordonmagnetic}
    -\chi_{1,2}'' + \left(\frac{(m+(Bu^2)/2)^2}{R^2 + u^2} \mp \frac{u(-2m + B (2R^2 + u^2))}{2(R^2 + u^2)^{3/2}}\right)\chi_{1,2} = \epsilon^2 \chi_{1,2}.
\end{equation}
In the eq.(\ref{kleingordonmagnetic}), the effective potential 
\begin{equation}
\label{magneticeffectivepotential}
U^{2}_{eff 1,2}=   \frac{(m+(Bu^2)/2)^2}{R^2 + u^2} \mp \frac{u(-2m + B (2R^2 + u^2))}{2(R^2 + u^2)^{3/2}}
\end{equation}
includes effects from the curved geometry, total angular momentum (spin and orbital), and the coupling to the magnetic field. Note that, for $u \gg R$ (outside the throat), the effective potential takes the form
\begin{equation}
    U^{2}_{eff}\approx \frac{m(m+1)}{u^2}+B\left(m-\frac{1}{2}\right)+\frac{B^2}{4}u^2,
\end{equation}
which is the effective potential for a $2+1$ massless Dirac fermion under an uniform magnetic field in a flat plane using cylindrical coordinates \cite{diracplanar,diracplanar2}. That is an expected result, since the graphene wormhole surface is asymptotic flat. Moreover, the effective potential in eq.(\ref{magneticeffectivepotential}) also exhibits the $m\rightarrow -m$ asymmetry. 

\begin{figure*} 
       \begin{minipage}[b]{0.46 \linewidth}
           \includegraphics[width=\linewidth]{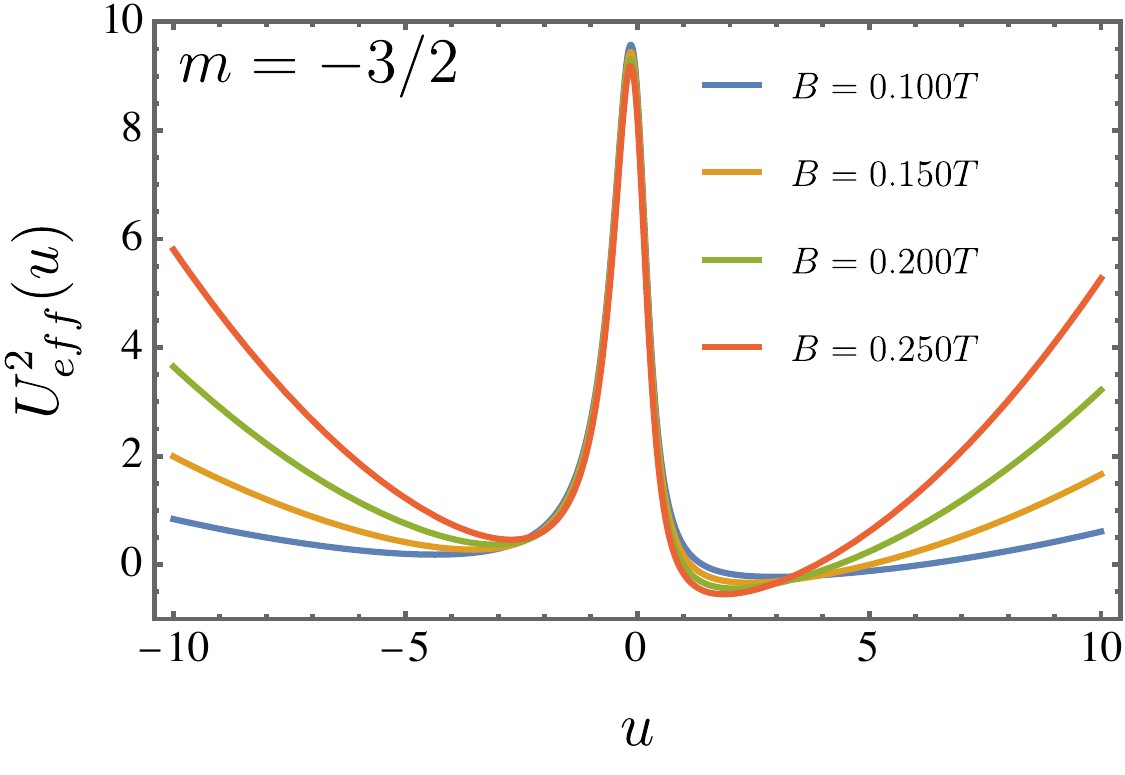}\\
           \caption{Effective squared potential for $m=-3/2$ for different values of $B$.}
          \label{U2}
       \end{minipage}\hfill
       \begin{minipage}[b]{0.47 \linewidth}
           \includegraphics[width=\linewidth]{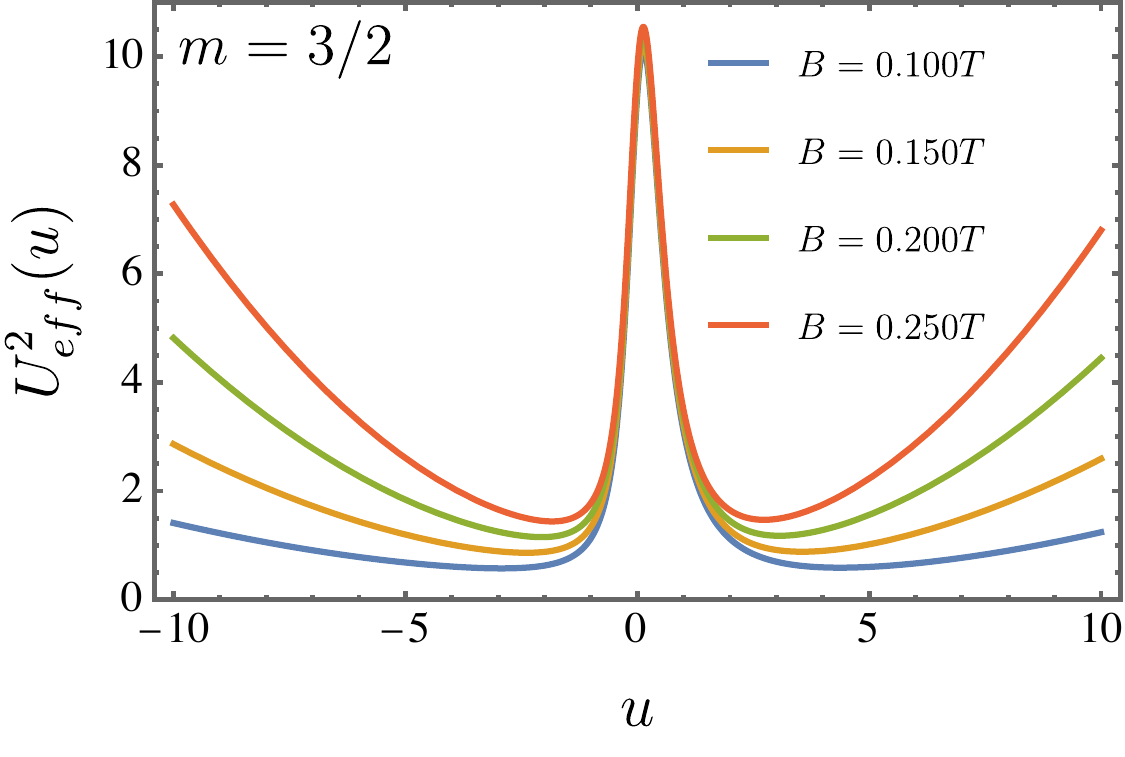}\\
           \caption{Effective squared potential for $m=3/2$.  Red line(B=0.1), blue line (B=0.15) and the green line (B=0.2).}
           \label{U3}
       \end{minipage}
\end{figure*}

Due to the complexity of eq.(\ref{kleingordonmagnetic}) we employ numerical methods to obtain the first excited states and their respective energy spectrum (Landau levels). In the figures (\ref{U2}) and (\ref{U3}) we plotted the effective potential $U^{2}_{eff}$ for $m=0$ and $m=1$, respectively.
Note that for $u\gg R$, the effective potential diverges as $\frac{B^2}{4}u^2$, whereas for $u\ll R$ the potential is dominated by the geometric and angular momentum terms (finite barrier for $m\neq 0$). 

\begin{figure*} 
       \begin{minipage}[b]{0.48 \linewidth}
           \includegraphics[width=\linewidth]{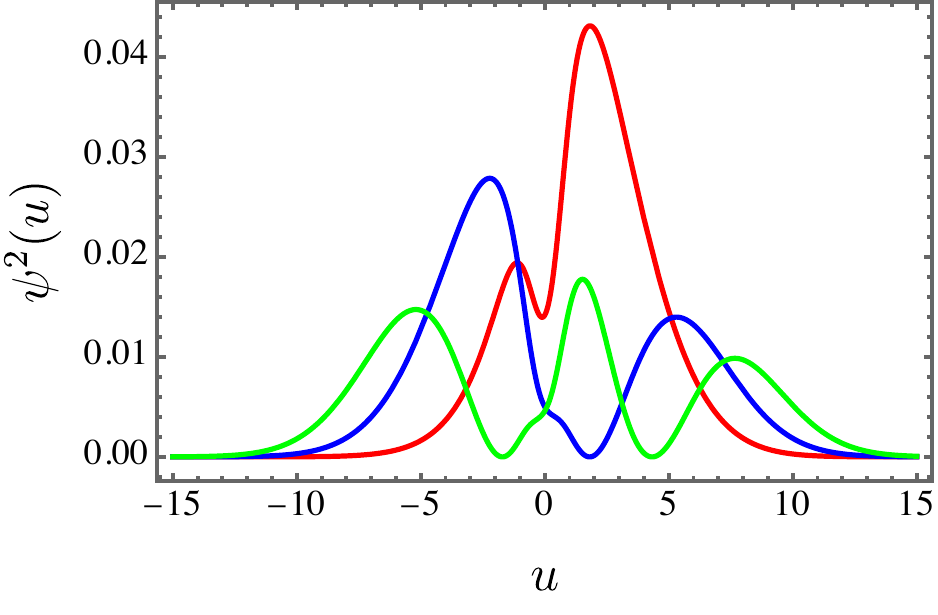}\\
           \caption{Density of states for $B=0.1$ and $m=1/2$. The ground state (red line) has a peak in the upper layer whereas the first excited state (blue line) is more localized in the lower layer. The second excited state exhibits three less distinct peaks.}
          \label{magm0}
       \end{minipage}\hfill
       \begin{minipage}[b]{0.48 \linewidth}
           \includegraphics[width=\linewidth]{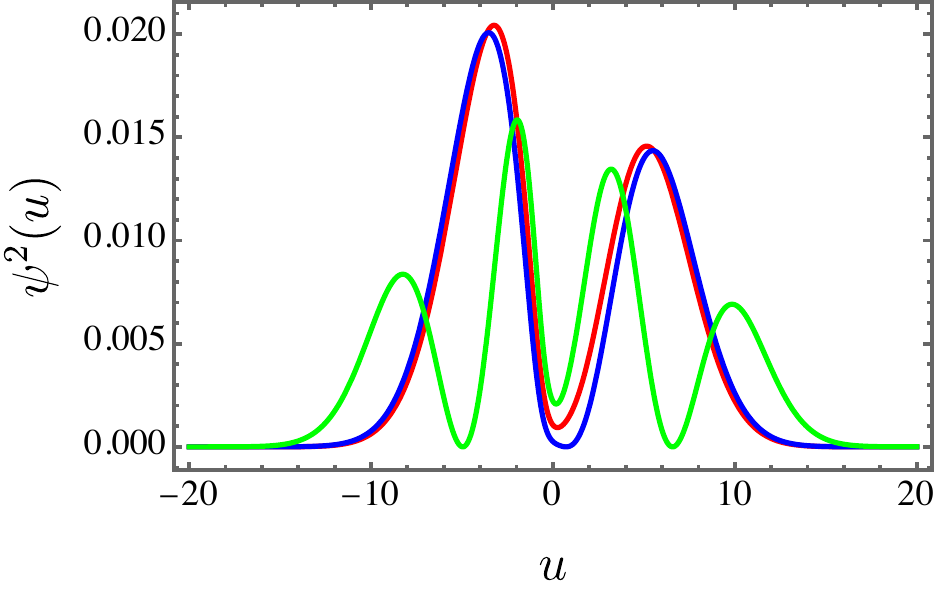}\\
           \caption{Density of states for $B=0.1$ and $m=3/2$.The ground state (red line) has a peak in the upper layer whereas the first excited state (blue line) is more localized in the lower layer. The second excited state exhibits three less distinct peaks.}
           \label{magm2}
       \end{minipage}
\end{figure*}

We plotted the first Landau levels for $eB=1$, $R=1$, $\beta=1$ and $m=0$ (s state) in the fig.(\ref{magm0}). Note that the first excited state (red line) is located on the upper layer, whereas the second (blue) and the third (green) have two asymmetric peaks around the origin. For  $eB=1$, $R=1$, $\beta=1$ and $m=2$ in the fig.(\ref{magm2}), the first excited state already has two asymmetric peaks displaced from the origin. Nonetheless, it is worthwhile to mention that the probability density  does not vanish at the origin. Note that for $u\gg R$, the wave function exhibits the usual exponential decay due to the external magnetic field \cite{diracplanar}.
Therefore, the external magnetic field allow us to confine the electron around the wormhole throat. However, due to the curved geometry and the strain, the electron is not symmetrically confined around the wormhole.

Finally, the ground state zero mode under the action of the magnetic field is modified by
\begin{eqnarray}
\label{magneticzeromode}
    \nonumber \psi^{0}_{1,2}(u)&=&(R^2+u^2)^{-1/4}e^{-\bar{\beta}\sigma(u)}e^{\mp m\sinh^{-1}(u/R)}\\
    &&\times e^{\mp \frac{eB}{4}(u\sqrt{R^2 +u^2} +R^2 \tanh^{-1}(u/\sqrt{R^2 +u^2}))}.
\end{eqnarray}
In the eq.(\ref{magneticzeromode}), the magnetic field introduces another exponential factor whose argument is a parity-odd function. Accordingly, the magnetic field enhances the chiral separation of the eletronic modes. However, note that changing the sign of the magnetic field for a given angular momentum number $m$, the magnetic field might reduce the chiral separation between the upper and lower layers.

\section{Final remarks and perspectives}
\label{section6}

We investigated the curvature, strain and magnetic field effects upon a massless relativistic electron on a graphene wormhole surface. The graphene wormhole was described by a catenoid surface which smoothly connects two asymptotic flat graphene planes (layers). In this sense, the geometry proposed was a smooth generalization of the graphene wormhole shown in the Ref.\cite{wormhole}. The effective Dirac Hamiltonian containing strain--dependent terms was obtained in Ref.\cite{vozmediano} and extended in Ref.\cite{vozmediano2}.

Due to the axial symmetry, we considered an isotropic strain tensor localized near the wormhole throat, similar to the behavior of the Gaussian curvature. Indeed, since the curved geometry of the throat was obtained due to deformation of the lattice structure, it was expected that both curvature and strain were localized around this region. It turned out that the pseudo--magnetic potential due to the strain $\Gamma_u$ had only components along the meridian coordinate $u$, whereas the spin connection $\Omega_\phi$ pointed along the parallel direction $\phi$. In this manner, despite having the same origin (the lattice deformation), these two interactions had distinct effects on the electron. Moreover, by applying the external magnetic field, a true magnetic potential $\vec{A}$ also acted on the electron. Nevertheless, although $\vec{A}$ only had the angular component $A_\phi$, the spin connection $\Omega_\phi$ was parity--odd, whereas $A_\phi$ was parity--even under the transformation $u\rightarrow -u$. 

The strain and spin--curvature breaks the parity invariance of the effective Hamiltonian. Moreover, the spin--connection term led to a chiral invariance $m\rightarrow -m$. By employing the supersymmetric quantum mechanical (QMSUSY) approach, we found that the strain vector yielded to an exponential suppression of the wave function, whereas the spin connection led to a power--law decay. In absence of magnetic field, the superpotential was given by the spin--curvature term which increased the amplitude of the probability density in the upper layer for $m>0$ and in the lower layer for $m<0$. A similar chiral behavior was found in graphene nanoribbons in a helical shape \cite{atanasov}.
Since the graphene structure is asymptotically flat, for a vanishing strain vector, the wave function behaves as a free wave in a flat plane \cite{watanabe}. The inclusion of a uniform magnetic field confined the electronic states near the wormhole throat. These Landau levels were modified by the spin--curvature and the strain interactions turned out to be an asymmetric probability distribution.

In addition, this work revealed that, despite the coupling of strain, curvature and magnetic field in the effective Dirac Hamiltonian being similar, they possessed rather different effects. As a result, we pointed out a couple of perspectives for further investigation. A noteworthy extension of the present work could consider the effects of dynamical strain (phonons) or corrugations on the graphene wormhole. Furthermore, the chiral breaking due to the spin--curvature coupling suggests an interesting spin--Hall current to be analyzed. Moreover, the analysis of the geometric Aharonov--Bohm like phase due to the concentrated curvature around the wormhole throat seems a worthy perspective as well.

\section*{Acknowledgments}
\hspace{0.5cm} J.E.G.Silva thanks the Conselho Nacional de Desenvolvimento Cient\'{\i}fico e Tecnol\'{o}gico (CNPq), grants n$\textsuperscript{\underline{\scriptsize o}}$ 304120/2021-9 for financial support. Particularly, A. A. Araújo Filho would like to thank Fundação de Apoio à Pesquisa do Estado da Paraíba (FAPESQ) and Conselho Nacional de Desenvolvimento Cientíıfico e Tecnológico (CNPq)  -- [200486/2022-5] and [150891/2023-7] for the financial support. Most of these calculations were performed by using the \textit{Mathematica} software.


\end{document}